\documentclass[aps, prd, superscriptaddress, twocolumn, nofootinbib]{revtex4-2}

\usepackage[colorlinks, linkcolor=red, anchorcolor=green, citecolor=blue]{hyperref}
\usepackage{amsmath}
\usepackage{amssymb}
\usepackage{graphicx}
\usepackage{color}
\usepackage{braket}
\usepackage{orcidlink}
\usepackage{comment}
\graphicspath{{figures}}

\newcommand{\bx}{\boldsymbol{x}}
\newcommand{\bp}{\boldsymbol{p}}
\newcommand{\bP}{\boldsymbol{P}}
\newcommand{\rP}{\mathrm{P}}
\newcommand{\bph}{\boldsymbol{\phi}}
\newcommand{\mW}{\mathcal{W}}
\newcommand{\mI}{\mathcal{I}}
\newcommand{\mS}{\mathcal{S}}
\newcommand{\mD}{\mathcal{D}}
\newcommand{\bQ}{\boldsymbol{Q}}
\newcommand{\basis}{\mathcal{P}}
\newcommand{\basisleft}{\mathcal{Q}}
\newcommand{\mass}{m_m}
\newcommand{\hatprod}{\mathop{\widehat{\prod}}}

\colorlet{conyorange}{-cyan!70!black}

\begin{document}

\title{
Spectral BBGKY: a scalable scheme for nonlinear Boltzmann and correlation kinetics
}

\author{Xingjian Lu}
\email[]{lus21@mails.tsinghua.edu.cn}
\affiliation{Department of Physics, Tsinghua University, Beijing 100084, China}

\author{Shuzhe Shi}
\email[]{shuzhe-shi@tsinghua.edu.cn}
\affiliation{Department of Physics, Tsinghua University, Beijing 100084, China}

\begin{abstract}
The Bogoliubov--Born--Green--Kirkwood--Yvon (BBGKY) hierarchy provides a time-reversal-symmetric framework for describing the nonequilibrium evolution of many-body systems. Despite the success of Boltzmann-based numerical approaches, systematically extending beyond this lowest-order truncation to the full nonlinear BBGKY hierarchy remains a major challenge. Moreover, even at the Boltzmann level, accurately treating the nonlinear collision term still presents significant difficulties. Here we propose the spectral BBGKY hierarchy, an analytically equivalent and numerically tractable reformulation of the conventional BBGKY hierarchy. The spectral formulation reduces the original $6n$-dimensional phase-space problem to the evolution of spectral coefficients over the $3n$-dimensional coordinate space, thereby eliminating the need for momentum-space discretization. In addition, we develop an analytic scheme for computing the collision integrals, which achieves high accuracy and removes the need for ensemble averaging over repeated stochastic evolutions from the same initial state. The scheme evaluates the full eight-dimensional integral exactly for massless particles, and reduces it to a three-dimensional one for massive particles. The validity of the spectral BBGKY hierarchy is verified through conservation law analysis, comparison with an analytical solution, convergence tests, and analysis of spectral coefficient leakage. At minimal truncation, the spectral BBGKY yields a spectral nonlinear Boltzmann equation that captures nonlinear dynamics with a computational cost comparable to that of conventional linearized approaches. When extended to higher-order truncations ($n \ge 2$), the spectral BBGKY hierarchy provides a flexible framework for studying multiparticle correlations across a wide range of systems---from ultracold atomic gases and quark-gluon plasma produced in relativistic heavy-ion collisions to the primordial plasma of the early universe. This framework advances our ability to investigate the early thermalization puzzle in relativistic heavy-ion collisions and to elucidate the applicability of hydrodynamics at remarkably early stages of quark-gluon plasma evolution.
\end{abstract}

\maketitle
\section{Introduction}

Non-equilibrium statistical mechanics is a challenging area of physics that has attracted increasing attention~\cite{
10.1093/oso/9780195140187.001.0001,
Calabrese:2006rx,
Seifert_2008,
2008EL.....8120003S,
annurev:/content/journals/10.1146/annurev-conmatphys-070909-104101,
Esposito:2008dpw,
10.1063/1.881363,
Le:2022ntg,
2013EPJST.217...43L}, driven by both experimental progress and theoretical advances.
A wide range of physical systems---from ultracold atoms~\cite{Langen:2015uhu,Cao:2010wa,Le:2022ntg} to high-energy nuclear collisions~\cite{
Berges:2004ce, Ryblewski:2012rr, Schenke:2012wb, Kurkela:2015qoa}---has made it possible to explore how complex systems evolve and relax when driven far from equilibrium. At the same time, new theoretical frameworks~\cite{Jarzynski:1996oqb, 2008PhLB..670..135G, Berges:2004ce, 2013EPJST.217...43L} have opened the door to a deeper understanding of (pre)thermalization and the emergence of macroscopic behavior in isolated or open systems.
Kinetic theory, as a central framework within non-equilibrium statistical mechanics, offers a microscopic statistical description that bridges particle interactions and emergent macroscopic behavior.
Both theoretical and experimental studies support the effectiveness of kinetic theory in describing non-equilibrium processes across a wide range of systems, including ultracold atomic gases~\cite{Langen:2015uhu, Cao:2010wa, Le:2022ntg, 2013EPJST.217...43L}, carrier quasiparticles in semiconductors~\cite{KOSINA200193, RevModPhys.55.645, 1962PhRv..126.2002S, 2000fct..book.....L, 1988PhRvB..38.9721F, 1987PhRvB..36.1570F}, the quark-gluon plasma produced in relativistic heavy-ion collisions~\cite{Wang:1991hta, Zhang:1999bd, Berges:2004ce, Lin:2004en, Xu:2004mz, Xu:2007aa, Xu:2007jv, El:2007vg, Fochler:2008ts, Xu:2008av, Xu:2010cq, Ryblewski:2012rr, Strickland:2018ayk, Kurkela:2015qoa, Ochsenfeld:2023wxz}, and the primordial plasma of the early universe~\cite{Ma_1995, 1996ApJ...469..437S}.

Specifically, in the context of high-energy heavy-ion collisions, kinetic theory plays a crucial role in addressing the long-standing puzzle of early thermalization. Shortly after the observation of the quark-gluon plasma (QGP), relativistic hydrodynamics was found to successfully describe its collective behavior, including phenomena such as elliptic and radial flow (see, e.g.,~\cite{Shen:2014vra, Schenke:2010rr, Karpenko:2013wva, vanderSchee:2013pia, Pang:2018zzo, Du:2019obx}). 
However, both macroscopic~\cite{1959flme.book.....L, Baier:2007ix} and microscopic~\cite{Muller:1967zza, Israel:1979wp, Denicol:2012cn} derivations of hydrodynamic equations rely on the assumption that the medium is close to thermal equilibrium. For the former, only lower-order terms in the gradient expansion are taken into account. For the latter, equation of motion is truncated up to linear non-equilibrium corrections in the phase space distribution. 
Thus, hydrodynamics presumes that the system has reached local thermal equilibrium, a condition typically justified only when the mean free path is much shorter than the time evolution scale, in natural units ($k_B = c = \hbar = 1$).
In heavy-ion collisions, however, this assumption becomes highly nontrivial. The onset time of hydrodynamic applicability is inferred to be approximately $1~\mathrm{fm}/c$~\cite{Nijs:2020roc, Nijs:2020ors}, which is comparable to the mean free path of the constituent particles.
This coincidence challenges the internal consistency of applying hydrodynamics at such early stages and highlights a gap in our understanding of how the system rapidly approaches equilibrium from an initially far-from-equilibrium state.
Kinetic theory provides a first-principles framework to explore this transition, offering microscopic insight into the emergence of fluid-like behavior.

The Liouville equation is the fundamental starting point of kinetic theory in statistical mechanics. It governs the exact time evolution of the full phase-space distribution function for an interacting system of $N$ particles.
In practice, solving the Liouville equation is intractable for realistic systems due to the exponential complexity associated with the full $6N$-dimensional phase space. 
To address this challenge, the Bogoliubov--Born--Green--Kirkwood--Yvon (BBGKY) hierarchy~\cite{bogoliubov1946kineticEnglish,bogoliubov1946kineticRussian,bogoliubov1947kinetic,1946JChPh..14..180K,1947JChPh..15...72K,1946RSPSA.188...10B,yvon1935theorie} was developed as a systematic framework for many-body dynamics. Instead of directly evolving the full $N$-body distribution function, the BBGKY hierarchy reformulates the problem in terms of reduced distribution functions, which describe single-particle distributions as well as two-body and higher-order correlations in a coupled manner. Although the equations remain hierarchically interconnected, this formulation provides a fundamental and systematic approach to modeling the nonequilibrium evolution of interacting many-body systems.
The well-known Boltzmann equation represents the lowest-order truncation of the BBGKY hierarchy. It effectively closes the hierarchy at the single-particle distribution level by invoking the molecular chaos assumption, which neglects correlations between particles prior to their collisions.
Due to its transparent physical interpretation, the Boltzmann equation has become a cornerstone of nonequilibrium statistical mechanics. It is widely employed to model transport processes, relaxation phenomena, and thermalization across a broad range of physical systems~\cite{Langen:2015uhu, Cao:2010wa, Le:2022ntg, 2013EPJST.217...43L, Berges:2004ce, Ryblewski:2012rr,  Kurkela:2015qoa, Ma_1995,1996ApJ...469..437S, KOSINA200193, RevModPhys.55.645, 1962PhRv..126.2002S, 2000fct..book.....L, 1988PhRvB..38.9721F, 1987PhRvB..36.1570F}.

However, it should be emphasized that the Boltzmann equation neglects correlations between particles prior to their collisions. In some physical systems, such correlations play an essential role in the system's dynamical evolution and must be retained for a faithful description.
For example, in the early stages of high-energy heavy-ion collisions, the initial state is dominated by strongly interacting particles exhibiting significant momentum and spatial correlations. This is particularly true for scenarios described by the minijet picture (energetic quarks and gluons produced in hard scatterings)~\cite{Wang:1991hta, Gyulassy:1994ew} or the Color Glass Condensate theory (a state governed by condensed non-Abelian gauge fields)~\cite{Schenke:2012wb}.
These initial correlations influence the subsequent development of radial flow, anisotropic flow, and other collective phenomena. Neglecting these correlations, as done in the standard Boltzmann approach, can lead to systematic inaccuracies in characterizing the early-time dynamics and thermalization rates. A related situation arises in plasmas with long-range Coulomb interactions, where two-particle correlations persist even at low densities due to collective screening effects and long-range tails in the interaction potential~\cite{2008PhyA..387..787C}.

Despite decades of progress, most existing numerical studies of the BBGKY hierarchy focus on the lowest-order truncation---namely, the Boltzmann equation---without incorporating explicit particle correlations. Even two-body correlations, the simplest form of multi-particle correlations, remain largely unexplored in numerical implementations. 
To the best of our knowledge, the most closely related approach is an analytical attempt within the relaxation time approximation~\cite{2025arXiv250100099G}, which replaces the full collision integral with a linearized relaxation term.
The common practice of truncating the BBGKY hierarchy at first order is largely due to the severe computational challenges involved in evolving $n$-particle reduced distribution functions. The governing equations are nonlinear integro-differential equations defined over a $6n$-dimensional phase space, making direct numerical simulation technically demanding and computationally expensive.

Existing numerical strategies for kinetic equations can be broadly categorized into two classes.
The first class consists of particle-based dynamical simulation methods. These algorithms simulate the microscopic dynamics of particles undergoing sequences of free streaming and collisions. Particle-based dynamical methods are further divided into geometrical and stochastic approaches. In geometrical methods, a collision is assumed to occur when two particles come sufficiently close to each other---closer than a threshold distance determined by the effective size of their interaction cross section~\cite{Geiger:1991nj, Molnar:2001ux, Molnar:2000jh, Zhang:1997ej, Sun:2022xjr}. Within this framework, the probability of a collision is binary---it is either unity or zero, depending solely on the spatial proximity of the particles. Since the cross section is interpreted as a physical area, this approach is referred to as the ``geometrical method.'' However, this method breaks Lorentz covariance and may result in acausal collisions, particularly when the mean free path is comparable to the range of interaction~\cite{Xu:2004mz}. 
In contrast, stochastic methods handle collisions based on transition rates, evaluating the probability of various microscopic processes occurring within a given spatial volume and time interval~\cite{Xu:2004mz}. A key advantage of this approach is that it naturally accommodates inelastic $2\leftrightarrow3$ scattering processes, which are difficult to model within the geometrical framework.
A representative implementation is the Boltzmann Approach of Multiparton Scatterings (BAMPS), developed by Xu and Greiner~\cite{Xu:2004mz}. BAMPS employs an extended stochastic method for the collision integral, enabling a fully covariant and consistent treatment of inelastic $2 \leftrightarrow 3$ back-reactions. The BAMPS was later extended to $2 \leftrightarrow 4$ collisions in~\cite{Sun:2022xjr}.
Both classes of particle-based dynamical methods---whether based on the geometrical or stochastic approach---are assisted by the test-particle method, in which the dynamics of a single physical particle are represented by a large number of test particles. 
While the rescaling of particle number and cross section in the test particle method preserves the correct evolution of the single-particle distribution, it disrupts the proper dynamics of multi-particle correlations. Consequently, particle-based dynamical methods fail to accurately describe the evolution of such correlations.
Moreover, despite improving the smoothness of the single-particle distribution function, the test-particle method still inevitably retains statistical fluctuations. 
To suppress these fluctuations and achieve reliable results, it is necessary to perform multiple simulations for a given initial condition and carry out ensemble averaging.

The second class of methods treats the reduced distribution functions as continuous fields and solves their evolution equations as a special type of partial differential equations (PDEs)---specifically, integro-differential equations. 
This class includes standard numerical solvers such as the finite difference method, finite volume method, and finite element method~\cite{AbraaoYork:2014hbk, Soudi:2021aar, Ochsenfeld:2023wxz}.
Due to the high dimensionality of phase space, solving the BBGKY hierarchy by regarding the reduced distribution functions as continuous fields becomes computationally intractable.
A widely adopted implementation is the Lattice Boltzmann Method (LBM)~\cite{Romatschke:2011hm, Mendoza:2009gm}. LBM is specifically designed to solve the linearized Boltzmann equation, with the momentum space discretized in such a way that the evolution of the energy-momentum tensor is preserved exactly. 

However, significant challenges remain in extending existing methods to nonlinear kinetic problems. First, for all PDE-based approaches, a major drawback is the need to discretize the full $6n$-dimensional phase space. As $n$ increases, both memory usage and computational cost scale exponentially, rendering even the $n = 1$ case nearly infeasible for practical nonlinear simulations.
Second, regardless of whether particle-based or PDE-based methods are employed, evaluating the collision term is computationally expensive. This is because it requires numerical integration at every time step and is subject to integration-induced uncertainties.
These challenges underscore the need for a scalable scheme for nonlinear Boltzmann and correlation kinetics---a goal that this work aims to achieve through the spectral BBGKY hierarchy.

The spectral method is a class of techniques for solving differential equations by expanding the solution in terms of a complete set of orthogonal basis functions. With appropriate basis choices, these methods are known for their high accuracy and low computational cost. In this work, we extend the spectral method to the BBGKY hierarchy by expanding the reduced distribution functions in terms of orthogonal basis functions defined over momentum space. This leads to the spectral BBGKY hierarchy---a reformulation that is analytically equivalent to, and numerically more tractable than, the conventional BBGKY hierarchy. A key advantage of this formulation is that it eliminates the need to discretize momentum space, thereby reducing the $6n$-dimensionality of the problem to $3n$---the number of spatial variables required to describe the evolution of the $n$-particle distribution function. A full description of the spectral BBGKY hierarchy can be found in Sec.~\ref{sec:spectral BBGKY hierarchy}.
Moreover, we develop an analytic scheme to compute the collision integrals for arbitrary collision kernels, as described in Sec.~\ref{sec:Collision Kernel}. 
The scheme evaluates the full eight-dimensional integral exactly for massless particles. For massive particles, it significantly simplifies the computation by reducing the original eight-dimensional integral to a three-dimensional one. 
This is achieved by further expanding the differential cross section in a separate set of orthogonal basis functions.
Finally, in Sec.~\ref{sec:Result}, we verify the proposed spectral BBGKY hierarchy and the analytic collision kernel through four evidences:
(i) exact preservation of conservation laws;
(ii) agreement with known analytical solutions~\cite{Bazow:2015dha};
(iii) systematic numerical convergence with respect to basis truncation; and
(iv) negligible leakage of expansion coefficients.
After presenting the summary and discussion in Sec.~\ref{sec:Discussion and Conclusion}, supplementary material is provided in Appendices~\ref{Appendix:Spherical Harmonics}-\ref{Appendix:Other Physical Quantity in Coefficients}, covering details on spherical harmonics, $3j$ symbols, and other relevant topics, respectively.

\subsection{Notations}
For convenience, we list the notations used in the main text.
\begin{itemize}
\item $k_B=c=\hbar=1$.
\item $\mass$: mass of the particle.
\item $\mathrm{i}$: the imaginary unit.
\item $\vartheta(x)$: the Heaviside step function.
\item $\lfloor x \rfloor$: the greatest integer not greater than $x$.
\item $Y_{\ell}^m$ and $Y_{\ell,m}$ denote the complex and real spherical harmonics, respectively.
\item 
We use the shorthand
\begin{align}
\int_{p_{\mu}} \equiv \int d (p_{\mu} u^{\mu} /\Lambda)\, d\Omega
\end{align}
for energy-angle integration in the local rest frame, and
\begin{align}
\int_{\bp} \equiv \int \frac{d^{3}\bp}{(2\pi)^3 p^0}
\end{align}
for the standard Lorentz-invariant phase-space measure with $p^0 = \sqrt{|\bp|^2 + \mass^2}$.
\item For clarity, throughout this paper, the indices $n, \ell, m$ are used solely as parameters of the expansion basis functions and do not follow the Einstein summation convention. Other indices, including but not limited to $i, j, k, s$, are used according to context and may or may not be subject to Einstein summation.
\item We use the standard abbreviation for the inner product of three-dimensional vectors, $\boldsymbol{a} \cdot \boldsymbol{b} \equiv \sum_{i\in\{x,y,z\}} a_i b_i$, and for four-dimensional vectors, $a \cdot b \equiv g^{\mu\nu} a_\mu b_\nu$, with the metric convention $g^{\mu\nu} = \mathrm{diag}(+1, -1, -1, -1)$.
\item The basis functions $\basis(p^\mu)$ used to expand the reduced distribution functions are defined identically for both four-momentum and three-momentum forms, i.e. $\basis(p^\mu) \equiv \basis(\bp)$.
\item The symbol $\hatprod$ denotes a compact notation for multiple nested sums or products, analogous to $\sum$ or $\prod$, but extended over multiple indices. For example,
\begin{align}\label{eq:hatprod}
\left[ \hatprod_{\chi = i,j} \sum_{k_\chi = 0}^{K} \right] =& \sum_{k_i = 0}^{K} \sum_{k_j = 0}^{K}\,,
\\
\left[ \hatprod_{\chi = i,j} \prod_{k_\chi = 0}^{K} \right] =& \prod_{k_i = 0}^{K} \prod_{k_j = 0}^{K}\,.
\end{align}
\item The multinomial coefficient, denoted as
\begin{align}\label{eq:multinomial}
\binom{n}{m_1, m_2, \ldots, m_k} = \frac{n!}{m_1! \, m_2! \, \cdots \, m_k!},
\end{align}
is defined for non-negative integers $m_1, m_2, \ldots, m_k$ such that $n = m_1 + m_2 + \cdots + m_k$.
\item We use the notation $n!!$ to denote the double factorial of $n$, defined as the product of all integers from $n$ down to $1$ that have the same parity as $n$. Specifically, 
\begin{align}\label{eq:doublefactorial}
    n!! = n(n-2)(n-4)\cdots
    \,,
\end{align}
with $0!! = 1$ and $(-1)!! = 1$ by convention.
\end{itemize}

\section{Spectral BBGKY Hierarchy}\label{sec:spectral BBGKY hierarchy}

In this work, we reduce the numerical complexity of the BBGKY hierarchy by expanding the momentum distribution in a complete and orthonormal basis, also referred to as a spectral expansion.
The spectral BBGKY hierarchy, which will be introduced in the present section, is an analytically equivalent alternative to the conventional BBGKY hierarchy. 
The spectral BBGKY hierarchy can completely replace the latter while significantly improving computational feasibility.
This formulation provides a rigorous and numerically tractable framework for the systematic analysis of nonlinear evolution in single-particle distributions and multi-particle correlations, laying the foundation for deeper investigations into their behavior in non-equilibrium systems.

\subsection{Conventional BBGKY Hierarchy}\label{subsection:The BBGKY hierarchy}
\emph{Classical BBGKY.} --- 
The BBGKY hierarchy~\cite{bogoliubov1946kineticEnglish,bogoliubov1946kineticRussian,bogoliubov1947kinetic,1946JChPh..14..180K,1947JChPh..15...72K,1946RSPSA.188...10B,yvon1935theorie} is a sequence of equations that describes the evolution of a multi-particle system in statistical mechanics. 
It is mathematically equivalent to Liouville's equation. 
For pedagogical purposes, we begin by reviewing the standard BBGKY hierarchy. 
Specifically, for a multi-particle system with interactions described by a two-body potential, the BBGKY hierarchy is derived as follows.

Consider a Hamiltonian dynamical system consisting of $N$ particles. 
Each particle in the system can be located in phase space, $\bph_i = (\bx_i, \bp_i)$, by its canonical coordinates $\bx_i$ and conjugate momenta $\bp_i$, where $i = 1, \ldots, N$.
The time evolution of the phase-space distribution
$
    \rP_{(N)} \equiv \rP_{(N)}(t,\bx_1,\dots,\bx_N,\bp_1,\dots,\bp_N)
$
is governed by the Liouville equation,
\begin{align}
\begin{split}
&
    \frac{\partial \rP_{(N)}}{\partial t} 
    +
    \sum_{i=1}^N \frac{\bp_i}{p_i^0}\cdot\frac{\partial \rP_{(N)}}{\partial \bx_i}
=
    \sum_{i=1}^N \sum_{\substack{j=1\\ j\neq i}}^{N} (\nabla_{\bx_i}\Phi_{ij}) \cdot \frac{\partial \rP_{(N)}}{\partial \bp_i} 
    \,,
\end{split}\label{eq:LiouvilleEquation}
\end{align}
where $\Phi_{ij} \equiv \Phi_{ij}(|\bx_i - \bx_j|)$ is the pair interaction potential between particles $i$ and $j$, and $p^0$ is the energy of a particle, satisfying $(p^0)^2 = |\bp|^2 + \mass^2$, where $\mass$ is the mass of the particle.

The BBGKY hierarchy focuses on the evolution of $n$-body reduced distribution functions, defined by integrating out the phase space of the $(N-n)$ particles not of interest,
\begin{align}
\begin{split}
    \rP_{(n)} 
\equiv\;
    \rP_{(n)}(t,\bph_1,\cdots,\bph_n) 
=\;
    \int \rP_{(N)} \prod_{j={n+1}}^{N}d^6\bph_j\,,
\end{split}
\label{eq:reduced_DF}
\end{align}
where $n \le N$ and $d^6 \bph = \frac{d^3 \bx\, d^3 \bp}{(2\pi)^3}$.
The kinetic equation obeyed by the $n$-body reduced distribution functions, is obtained by integrating Eq.~\eqref{eq:LiouvilleEquation} with respect to $\prod_{j={n+1}}^{N}d^6\bph_j$,
\begin{align}\label{eq:standardBBGKY}
\begin{split}
&
    \frac{\partial \rP_{(n)}}{\partial t} 
    + \sum_{i=1}^n \frac{\bp_i}{p^0}\cdot\frac{\partial \rP_{(n)}}{\partial \bx_i}
    - \sum_{i=1}^n \sum_{\substack{j=1\\ j\neq i}}^{n} (\nabla_{\bx_i}\Phi_{ij}) \cdot \frac{\partial \rP_{(n)}}{\partial \bp_i}
\\&=\;
    (N-n)\sum_{i=1}^n
    \int d^6\bph_{n+1} (\nabla_{\bx_i}\Phi_{i,n+1}) \cdot \frac{\partial \rP_{(n+1)}}{\partial \bp_i}\,.
\end{split}
\end{align}
Eq.~\eqref{eq:standardBBGKY} thus represents the standard form of the BBGKY hierarchy.

\emph{Quantum BBGKY.} --- 
The standard BBGKY hierarchy given in Eq.~\eqref{eq:standardBBGKY} is classical: it treats the position and momentum of each particle as definite quantities and describes the interactions via a classical potential field.
In quantum theories, however, Heisenberg’s uncertainty principle prohibits the simultaneous precise knowledge of both the position and momentum of a particle, and interactions are described by the exchange of virtual bosons.

Taking quantum effects into account, the Wigner function---the Fourier transform of the two-point correlator of field operators---can be interpreted as the single-particle phase-space distribution.
Likewise, one may define the multi-particle distribution as the Fourier transform of multi-point correlators.
Indeed, the BBGKY hierarchy equations for generic or specific quantum mechanical or quantum field theories can be derived in a first-principles manner~\cite{2003PhLA..310..377K, 2010NCimC..33a..71B, MENDELSON2020107054, 2023arXiv231010940U, 2001nucl.th..10018L}.

In this work, we adopt a framework suitable for numerical computations by making a minimal generalization of the classical BBGKY hierarchy.
We consider an $N$-body system and treat the $N$-particle configuration $\{\bph\} \equiv \{\bph_i\}_{i\in[1,N]}$ in phase space as a microscopic state.
The evolution of the probability density of a phase-space configuration, $P_{(N)}(\{\bph\})$, is then governed by the Liouville equation,
\begin{align}
\begin{split}
&
    \frac{\partial \rP_{(N)}(\{\bph\})}{\partial t} 
    + \sum_{i=1}^N \frac{\bp_i}{p_i^0}\cdot\frac{\partial \rP_{(N)}(\{\bph\})}{\partial \bx_i} 
=\,
\\& 
    \int d^{6N}\{\bph'\} \Gamma_{\{\bph\}\to\{\bph'\}} 
    \Big(\rP_{(N)}(\{\bph'\}) - \rP_{(N)}(\{\bph\})\Big)
    \;.
\end{split}\label{eq:QuatumLiouvilleEquation}
\end{align}
Here, the transition rate $\Gamma_{\{\bph\}\to\{\bph'\}}$ denotes the probability per unit time for a configuration transition $\{\bph\}\to\{\bph'\}$. Owing to microscopic detailed balance, it satisfies $\Gamma_{\{\bph\} \to \{\bph'\}} = \Gamma_{\{\bph'\} \to \{\bph\}}$.

We then study the evolution of $n$-body ($n \leq N$) reduced distribution functions, as defined in Eq.~\eqref{eq:reduced_DF}. In this work, we consider particles interacting via short-range forces and restrict attention to two-to-two elastic scattering processes, so that the transition rate $\Gamma_{\{\bph\} \to \{\bph'\}}$ is given by the superposition of all binary interactions,
\begin{align}
\begin{split}\label{eq:short_interaction}
    &\Gamma_{\{\bph\}\to\{\bph'\}} 
\\=\;&
    \sum_{\substack{i,j=1\\i < j}}^{N}
    \frac{W_{(p_i^\mu,p_j^\mu \to {p'}_{i}^{\mu},{p'}_{j}^{\mu})}}{p_i^0\, p_j^0\, {p'}_{i}^0\, {p'}_{j}^0}
    \bigg(\prod_{k\neq i,j} \delta^{(6)}(\bph_k-\bph'_k)\bigg)
\\&\qquad\times
    \delta^{(3)}(\bx_i-\bx_j)
    \delta^{(3)}(\bx'_i-\bx'_j)
    \delta^{(3)}(\bx_i-\bx'_i)
\\&\qquad\times
    (2\pi)^4
    \delta^{(4)}(p_i^\mu + p_j^\mu - {p'}_{i}^{\mu} - {p'}_{j}^{\mu})
    \,.
\end{split}
\end{align}
Although only short-range interactions are considered in this study, long-range effects can be accounted for by incorporating gauge fields.
For convenience, we define a more compact form of the transition rate, which already incorporates energy-momentum conservation,
\begin{align}\label{eq:W}
\begin{split}
&
    \mW_{(\bp_1,\bp_2 \rightarrow \bp_{3},\bp_{4})}
\\=\;&
    W_{(p_1^\mu,p_2^\mu \to p_{3}^{\mu},p_{4}^{\mu})}
    (2\pi)^4
    \delta^{(4)}(p_1^\mu + p_2^\mu - p_{3}^{\mu} - p_{4}^{\mu})
    \;.
\end{split}
\end{align}
The kinetic equation obeyed by the $n$-body reduced distribution function is then obtained by integrating Eq.~\eqref{eq:QuatumLiouvilleEquation} with respect to $\prod_{j={n+1}}^{N}d^6\bph_j$,
\begin{align}\label{eq:QuantumBBGKY}
\begin{split}
    &\frac{\partial \rP_{(n)}(\{\bph\})}{\partial t} 
    + \sum_{i=1}^n \frac{\bp_i}{p_i^0}\cdot\frac{\partial \rP_{(n)}(\{\bph\})}{\partial \bx_i}
\\=\;&
    \sum_{\substack{i,j=1\\i < j}}^{n}
    \delta^{(3)}(\bx_i-\bx_j)
    \frac{1}{p_i^0 p_j^0}
    \int_{\bp'_i,\bp'_j}
    \mW_{(\bp_i,\bp_j\to \bp_i',\bp_j')}
\\&\times
    \big(\rP_{(n)}(\bp_i', \bp_j', \cdots) - \rP_{(n)}(\bp_i, \bp_j, \cdots)\big)
\\+\;&
    (N-n)\sum_{i=1}^{n}
    \frac{1}{p_i^0}
    \int_{\bp'_i, \bp'_j, \bp_j}
    \mW_{(\bp_i,\bp_j\to \bp_i',\bp_j')}
\\&\times
    \big(\rP_{(n+1)}(\bp_i', \cdots; \bp_j') - \rP_{(n+1)}(\bp_i, \cdots; \bp_j)\big)\,,
\end{split}
\end{align}
where $\int_{\bp} = \int \frac{d^{3}\bp}{(2\pi)^3 p^0}$.

\subsection{Spectral Reformulation of BBGKY}\label{subsection:spectral BBGKY Hierarchy}

The numerical challenges associated with reduced distribution functions in the conventional BBGKY hierarchy motivate us to seek a more efficient approach.
In this work, we address this issue by expanding the momentum dependence of the distribution function in terms of a set of orthogonal and complete basis functions.
This transforms both the distribution functions and their evolution equations into an equivalent formulation in terms of expansion coefficients.
Such an approach is commonly referred to as the \textit{spectral method} in the literature~\cite{Press2007}.

By choosing an expansion basis that is well adapted to the underlying physics, one can capture the essential features of the system by focusing on a few of the lowest-order expansion coefficients.
We expand the distribution functions in a special frame $S$ (defined below) using a set of basis functions that are both orthogonal and complete,
\begin{align}\label{eq:basis}  
\begin{split}
&
    \basis_{n,\ell,m}(p_{\mu})  
\\=\;&
    e^{-p_{\mu} u^\mu / \Lambda}  
    \left( \frac{p_{\mu} u^\mu}{\Lambda} \right)^\ell  
    Y_{\ell,m}(\theta,\phi)  
    L_n^{(2\ell+2)}\left( \frac{p_{\mu} u^\mu}{\Lambda} \right)  
    \,,  
\end{split}
\end{align}
where
$p_\mu$ is the four-momentum;
$u^\mu$ is the four-velocity of the special frame $S$ relative to the lab frame, satisfying $u^\mu u_\mu = 1$;
$\Lambda$ is a parameter with the dimensions of energy (not necessarily the equilibrium temperature of the system);
$Y_{\ell,m}(\theta,\phi)$ are the real spherical harmonics;
$\theta$ and $\phi$ are the polar and azimuthal angles of the three-momentum in spherical coordinates defined in frame $S$;
$L_n^{(2\ell+2)}$ are the associated Laguerre polynomials;
and the metric is $g^{\mu\nu} = \mathrm{diag}(+1,-1,-1,-1)$.
The conventions for spherical harmonics used in this work are summarized in Appendix~\ref{Appendix:Spherical Harmonics}. The special frame $S$ is defined such that the total momentum of the system of interest vanishes in this frame. If the system of interest is a fluid cell that is sufficiently small compared to the macroscopic scale but much larger than the microscopic scale, $u^\mu$ is the fluid four-velocity and frame $S$ is the local rest frame.

We note that the choice of real spherical harmonics has a notable impact on both numerical efficiency and computational cost.
Using real spherical harmonics ensures that the expansion coefficients remain real, as the distribution functions themselves are necessarily real-valued.
In contrast, employing complex spherical harmonics results in complex expansion coefficients, which leads to increased computational overhead.

The basis functions in Eq.~\eqref{eq:basis} satisfy the orthogonality condition,
\begin{align}  
    \int_{p_{\mu}}
    \basis_{n,\ell,m}(p_{\mu}) 
    \basisleft_{n',\ell',m'}(p_{\mu})
    = \delta_{n,n'} \delta_{\ell,\ell'} \delta_{m,m'} \,,  
\end{align}  
where the dual basis functions are defined as
\begin{align}\label{eq:dual_basis} 
\begin{split}  
    \basisleft_{n,\ell,m}(p_{\mu})  
=\;&
    \frac{n!}{(2\ell+n+2)!}  
    \left( \frac{p_{\mu} u^\mu}{\Lambda} \right)^{\ell+2}  
\\&\times
    Y_{\ell,m}(\theta,\phi)  
    L_n^{(2\ell+2)}\left( \frac{p_{\mu} u^\mu}{\Lambda} \right)  
    \,,
\end{split}  
\end{align}
and $\int_{p_{\mu}} = \int d (p_{\mu} u^{\mu} /\Lambda), d\Omega$.
For the sake of neatness, we flatten the $(n,\ell,m)$ indices in the basis function~\eqref{eq:basis} into a single index, $i$. This naturally leads to the Kronecker delta $\delta_{i,i'} = \delta_{n,n'}\delta_{\ell,\ell'}\delta_{m,m'}$ and the summations
$
    \sum_{i}  
    = \sum_{\ell=0}^{\infty} \sum_{m=-\ell}^{\ell} \sum_{n=0}^{\infty}
$.
We expand the reduced distribution functions $\rP_{(n)}$ as follows,
\begin{align}\label{eq:Pn_decompose}
\begin{split}
    \rP_{(n)}(\{\bph\})
=\;&
    \basis_{i_1} (\bp_1)
    \basis_{i_2} (\bp_2)
    \cdots
    \basis_{i_n} (\bp_n)
\\&\times
    \rP^{i_1 i_2 \cdots i_n}(t, \bx_1, \bx_2, \cdots, \bx_n)  
    \,,  
\end{split}
\end{align} 
where $\basis(p^\mu) \equiv \basis (\bp)$ and the Einstein summation convention is implicitly applied to repeated indices $i_1, \cdots, i_n$. The coefficients $\rP^{i_1 i_2 \cdots i_n}(t, \bx_1, \bx_2, \cdots, \bx_n)$ encode the space-time dependence of the respective distributions.
For clarity, throughout this paper, the indices $n, \ell, m$ are used solely as parameters of a given basis function and do not follow the Einstein summation convention. Other indices, including but not limited to $i, j, k, s$, are used according to context and may or may not be subject to Einstein summation.

By substituting the spectral decomposition of the distribution functions~\eqref{eq:Pn_decompose} into the standard BBGKY hierarchy~\eqref{eq:QuantumBBGKY}, we derive the spectral BBGKY hierarchy~\eqref{eq:spectralBBGKY}. The time evolution of the expansion coefficients $\rP$ corresponding to the reduced $n$-particle distribution functions is governed by
\begin{align}
\begin{split}
&
    \frac{\partial}{\partial t} 
    \rP^{i_1 i_2 \cdots i_n}(t, \bx_1, \bx_2, \cdots, \bx_n)  
\\+\;&
    \sum_{a=1}^n 
    \boldsymbol{B}_{i_a j_a}
    \cdot\frac{\partial }{\partial \bx_a}
    \rP^{i_1 i_2 \cdots j_a \cdots i_n}(t, \bx_1, \bx_2, \cdots, \bx_n)   
\\=\;& 
    \sum_{\substack{a,b=1\\a < b}}^{n}
    \delta^{(3)}(\bx_a-\bx_b)
    C_{i_a i_b j_a j_b}
\\&\times
    \rP^{i_1 i_2 \cdots j_a \cdots j_b \cdots i_n}(t, \bx_1, \bx_2, \cdots, \bx_n)   
\\+\;&
    (N-n)\sum_{a=1}^{n}
    A_{i_a j_a j_b}
\\&\times
    \rP^{i_1 i_2 \cdots j_a \cdots i_n j_b}(t, \bx_1, \bx_2, \cdots, \bx_a , \cdots, \bx_n, \bx_a)\,.
\end{split}\label{eq:spectralBBGKY}
\end{align}
Here, we require that all parameters in the expansion basis, $\Lambda$ and $u^\mu$, are space-time independent. $\boldsymbol{B}_{ij}$ represents the free-streaming integrals in the kinetic equation, defined as
\begin{align}\label{eq:Free streaming Integral}
\begin{split}
    \boldsymbol{B}_{ij}
=\;&
    \int_{p_\mu} 
    \basisleft_{i}(p_{\mu})
    \basis_{j}(p_{\mu})
    \frac{\bp}{p^0} 
    \;.
\end{split}
\end{align}
The three-index tensor $A_{ijk}$ and four-index tensor $C_{ijks}$ on the right-hand side are collision kernels, defined as the differences between the gain and loss terms,
\begin{align}
\label{eq:A_original}
A_{ijk} =\;& A^{\mathrm{gain}}_{ijk} - A^{\mathrm{loss}}_{ijk} \,,\\
\label{eq:C_original}
C_{ijks} =\;& C^{\mathrm{gain}}_{ijks} - C^{\mathrm{loss}}_{ijks} \,,
\end{align}  
with the gain and loss terms given by
\begin{align}\label{eq:A_Gain}
\begin{split}
    A^{\mathrm{gain}}_{ijk}
=\;&
    \int_{p_{1\mu}}
    \int_{\bp_2,\bp_{3},\bp_{4}}
    \basisleft_{i}(p_{1\mu})
\\&\times
    \frac{1}{p_1^0}
    \mW_{(\bp_1,\bp_2 \rightarrow \bp_{3},\bp_{4})}
    \basis_j(p_{3\mu})
    \basis_k(p_{4\mu})
    \;,
\end{split}\\
\label{eq:A_Loss}
\begin{split}
    A^{\mathrm{loss}}_{ijk}
=\;&
    \int_{p_{1\mu}}
    \int_{\bp_2,\bp_{3},\bp_{4}}
    \basisleft_{i}(p_{1\mu})
\\&\times
    \frac{1}{p_1^0}
    \mW_{(\bp_1,\bp_2 \rightarrow \bp_{3},\bp_{4})}
    \basis_j(p_{1\mu})
    \basis_k(p_{2\mu})
    \;,
\end{split}\\
\label{eq:C_Gain}
\begin{split}
    C^{\mathrm{gain}}_{ijks}
=\;&
    \int_{p_{1\mu},p_{2\mu}}
    \int_{\bp_{3},\bp_{4}}
    \basisleft_{i}(p_{1\mu})
    \basisleft_{j}(p_{2\mu})
\\&\times
    \frac{1}{p_1^0 p_2^0}
    \mW_{(\bp_1,\bp_2 \rightarrow \bp_{3},\bp_{4})}
    \basis_k(p_{3\mu})
    \basis_s(p_{4\mu})\;,
\end{split}\\
\label{eq:C_Loss}
\begin{split}
    C^{\mathrm{loss}}_{ijks}
=\;&
    \int_{p_{1\mu},p_{2\mu}}
    \int_{\bp_{3},\bp_{4}}
    \basisleft_{i}(p_{1\mu})
    \basisleft_{j}(p_{2\mu})
\\&\times
    \frac{1}{p_1^0 p_2^0}
    \mW_{(\bp_1,\bp_2 \rightarrow \bp_{3},\bp_{4})}
    \basis_k(p_{1\mu})
    \basis_s(p_{2\mu})\,.
\end{split}
\end{align}

Expanding the distribution functions using the basis functions in Eq.~\eqref{eq:basis} offers several advantages, as listed below.

First, it is \textit{simple in Maxwell--Boltzmann equilibrium}.
The single-particle distribution function, $f(t,\bx,\bp)\equiv N P_{(1)}(t,\bx,\bp)$, can be decomposed as $f(t,\bx,\bp) = \basis_i f^i(t,\bx)$.
In general, when the system reaches single-particle equilibrium (the Maxwell--Boltzmann distribution), the distribution function takes the form $f_{\mathrm{eq}} = e^{-p_{\mu} u^\mu / T}$,
indicating that the system becomes isotropic. 
As a result, only the coefficient $f^{(n,0,0)}$ remains nonzero, given by
\begin{align}
    f^{(n,0,0)} =  
    2 \sqrt{\pi}
    \left(\frac{T}{\Lambda}\right)^{3}  
    \left(1-\frac{T}{\Lambda}\right)^n  
    \,. \label{eq:coef_thermal}
\end{align}
More specifically, if the energy scale parameter $\Lambda$ matches the equilibrium temperature of the system, i.e., $\Lambda = T$, then at equilibrium, only the coefficient $f^{(0,0,0)}$ remains nonzero,
\begin{align}\label{eq:coef_thermal_special}
    f^{(n,0,0)} = 2 \sqrt{\pi}\, \delta_{n,0}
    \,.
\end{align}
In numerical simulations, the expansion basis must be truncated.
The spectral method is applicable only when the expansion of the distribution function can be truncated; that is, $f^{(n,0,0)}$ in a single-particle equilibrium should decrease with increasing $n$, which is valid provided the condition $\frac{T}{\Lambda} < 2$ is satisfied.
This condition is easily met, as the equilibrium temperature $T_{\mathrm{eq}}$ of a multi-particle system can be determined by the Landau matching condition, even when the system has not yet reached equilibrium. 
The Landau matching condition ensures that the energy density computed from the distribution function $f$ matches that obtained from the equilibrium distribution $f_{\mathrm{eq}}$.

Second, \textit{in the ultra-relativistic regime where $\mass = 0$, a direct relation exists between the conserved quantities and the spectral coefficients.}
The particle number density, $J^t = \int_{\bp} p^0 f(t,\bx,\bp)$, and the four-momentum density, $T^{t\mu} = \int_{\bp} p^0 p^\mu f(t,\bx,\bp)$, depend only on one or two expansion coefficients (with $p^0$ being the temporal component of the four-momentum $p^\mu$),
\begin{align}
J^t =\;& \frac{4 \sqrt{\pi}}{(2\pi)^3} \Lambda^3 f^{(0,0,0)}\,,\label{eq:conservationAndCoefficient_1}
\\
T^{tt} =\;& \frac{12 \sqrt{\pi}}{(2\pi)^3} \Lambda^4 \big(f^{(0,0,0)} - f^{(1,0,0)}\big)\,,
\\
T^{t\mu} =\;& \frac{16 \sqrt{3\pi}}{(2\pi)^3} \Lambda^4 f^{(0,1,m_\mu)}
\,,\label{eq:conservationAndCoefficient_5}
\end{align}
where $m_x = +1$, $m_y = -1$, and $m_z = 0$.
Furthermore, as will be demonstrated later, the conservation laws for particle number, energy, and momentum can be satisfied exactly.

Third, it \textit{requires low memory and achieves high numerical accuracy}.
Traditional methods for solving integro-differential equations, such as finite difference and finite volume schemes, operate on discrete grids.
They approximate derivatives using finite differences and replace integrals with numerical summations.
The accuracy of these methods scales algebraically with grid spacing. For instance, a $k$-th order central difference yields an error of $O(\Delta_p^k)$, where $\Delta_p$ is the grid size.
In typical applications, second-order accuracy $O(\Delta_{p}^2)$ is used in momentum space, which can limit precision unless very fine grids are employed.
In contrast, spectral methods can exhibit exponential convergence when applied to sufficiently smooth problems with an appropriately chosen basis, with the error typically scaling as $O(e^{-M})$, where $M$ denotes the number of basis functions~\cite{Press2007}.
This enables spectral methods to attain significantly higher accuracy even with substantially lower memory usage.

Fourth, the spectral approach adopted in this work \textit{saves computation time} by allowing precomputation of collision integrals.
In traditional methods---whether based on partial differential equations (PDEs) or particle-based dynamics---the collision integral must be computed at every time step, which is computationally expensive.
In contrast, in the spectral BBGKY hierarchy, the collision kernel can be precomputed.
This substantially reduces computational cost and accelerates the time-evolution procedure.

\begin{figure}[!hbtp]
    \centering
    \includegraphics[width=1\linewidth]{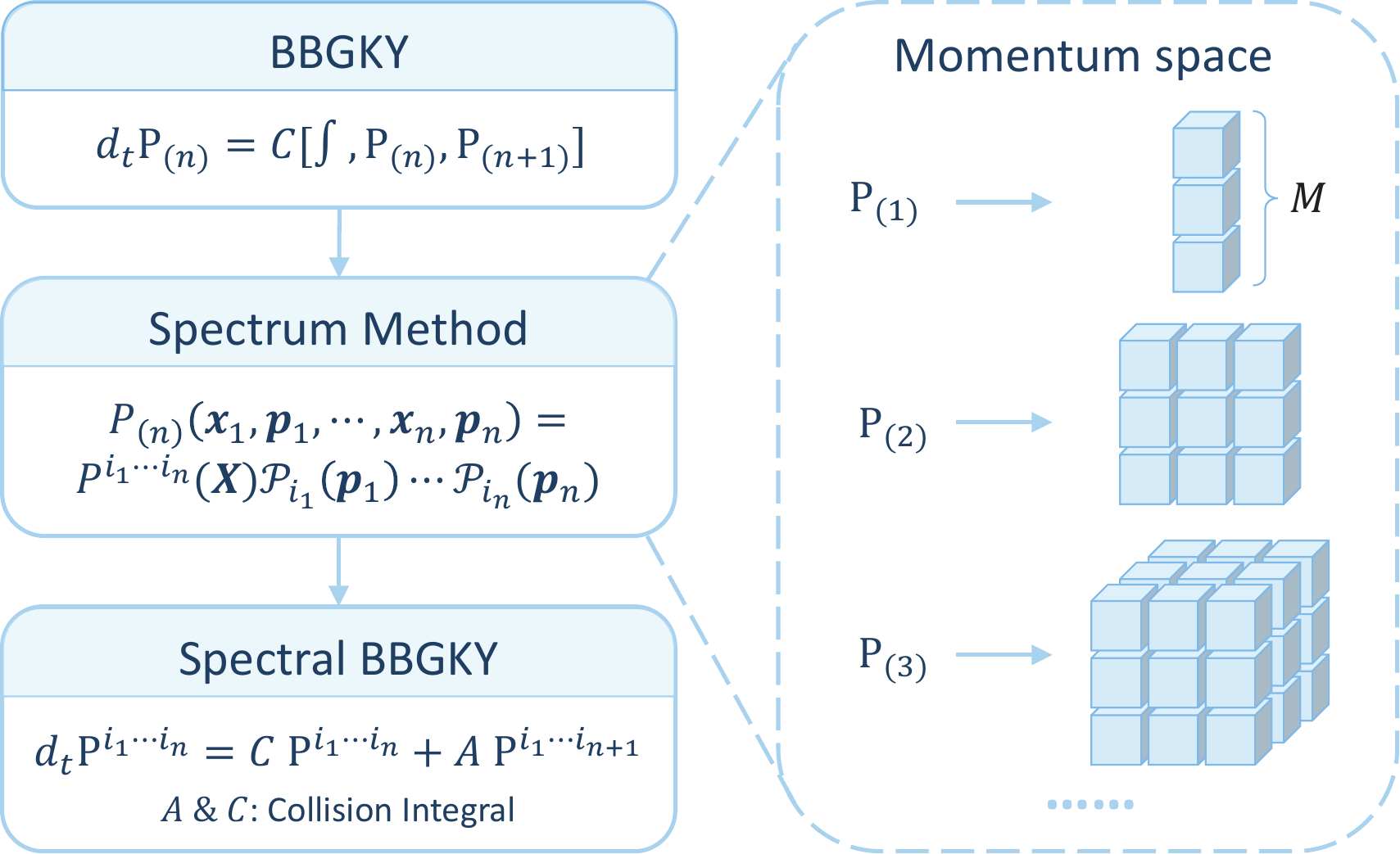}
    \caption{\textbf{Schematic overview of the spectral BBGKY hierarchy formulation.} A typical choice of truncation parameters is $(n_{\mathrm{max}}, \ell_{\mathrm{max}}) = (2, 2)$, which results in $M = (n_{\mathrm{max}} + 1)(\ell_{\mathrm{max}} + 1)^2 = 27$ basis functions used to encode the momentum dependence of the single-particle distribution. This truncation level has been verified to be sufficient for capturing the nonlinear evolution of key observables, including the energy-momentum tensor and moments.}
    \label{fig:schematic}
\end{figure}

The schematic of the spectral BBGKY hierarchy is summarized in Fig. \ref{fig:schematic}.
\section{Collision Kernel}\label{sec:Collision Kernel}

\subsection{Minimizing Collision Kernel Evaluations}\label{section:ApplyRatio}

\vspace{3mm}

The three- and four-index collision tensors ($A_{ijk}$ and $C_{ijks}$) in the spectral BBGKY hierarchy, i.e., Eqs.~(\ref{eq:A_Gain}-\ref{eq:C_Loss}), pose a substantial computational burden.
An expansion using $M$ basis functions requires the evaluation of $2\,M^3$ integrals for $A_{ijk}$ and $2\,M^4$ for $C_{ijks}$.
As an illustrative example, consider a typical basis configuration with expansion indices $n \in \{0,1,2\}$ and $\ell \in \{0,1,2\}$,
 which leads to $M = 27$ and $2 \times 27^4 \approx 1.1 \times 10^6$ required integrals for $C_{ijks}$ alone.
Due to the large number of required integrals, their computation becomes extremely time-consuming.

Fortunately, by exploiting the symmetry properties of spherical harmonics, the number of required collision integrals in the spectral BBGKY hierarchy can be significantly reduced.
With the reduction strategy presented in this subsection, the number of required integrals in the aforementioned example is reduced to 7,938---approximately $0.75\%$ of the original count.
This reduction becomes increasingly significant as $M$, particularly the upper limit of the expansion index $\ell$, increases.

We start by introducing unified tensor forms for the collision integrals appearing in the spectral BBGKY hierarchy. Specifically, we define higher-rank tensors $\tilde{A}_{ijksq}$ and $\tilde{C}_{ijksqw}$, which consolidate the gain and loss terms into single, unified expressions.
The five-index tensor $\tilde{A}_{ijksq}$ is defined as
\begin{align}\label{eq:A_5index_original}
\begin{split}
    \tilde{A}_{ijksq}
=\;&
    \int_{p_{1\mu}}
    \int_{\bp_2,\bp_{3},\bp_{4}}
    \tilde{\basisleft}_{i}^*(p_{1\mu})
    \frac{1}{p_1^0}
    \mW_{(\bp_1,\bp_2 \rightarrow \bp_{3},\bp_{4})}
\\&
    \times
    e^{\mathrm{s}_\mathrm{Q} \frac{P^\mu u_\mu}{\Lambda}}
    \tilde{\basis}_j(p_{1\mu})
    \tilde{\basis}_k(p_{2\mu})
    \tilde{\basis}_s(p_{3\mu})
    \tilde{\basis}_q(p_{4\mu})
    \,,
\end{split}
\end{align}
$\tilde{\basis}$ and $\tilde{\basisleft}$ are, respectively, the (dual) basis functions redefined by replacing the real spherical harmonics in Eqs.~\eqref{eq:basis} and~\eqref{eq:dual_basis} with the complex ones.
In the basis expansion, real spherical harmonics are used to ensure that the expansion coefficients remain real, thereby reducing the computational cost during time evolution. 
In contrast, here and throughout this section, we employ complex spherical harmonics in order to fully exploit their well-established mathematical properties and symmetries.
With the relation between real and complex spherical harmonics given in Appendix~\ref{Appendix:Spherical Harmonics}, one may formally express the transformations between them as $\basis_j = \mathcal{Y}_{jj'}\tilde{\basis}_{j'}$ and $\basisleft_j = \mathcal{Y}_{jj'}^*\tilde{\basisleft}_{j'}$. The standard expressions for $\mathcal{Y}_{jj'}$, $\tilde{\basis}$, and $\tilde{\basisleft}$ are defined in Eqs.~\eqref{eq:SH_real_vs_complex}, \eqref{eq:basis_complex}, and \eqref{eq:dual_basis_complex}, respectively.

The gain and loss terms defined in Eqs.~\eqref{eq:A_Gain} and~\eqref{eq:A_Loss} are related to $\tilde{A}_{ijksq}$ through the following identities,
\begin{align}
    A^{\text{gain}}_{ijk} =\;& 4\pi\,  \mathcal{Y}_{ii'}^* \mathcal{Y}_{jj'} \mathcal{Y}_{kk'} 
\tilde{A}_{i'00j'k'} \,, \label{eq:Again_A5index}\\
    A^{\text{loss}}_{ijk} =\;& 4\pi\,  \mathcal{Y}_{ii'}^* \mathcal{Y}_{jj'} \mathcal{Y}_{kk'} 
    \tilde{A}_{i'j'k'00} \,, \label{eq:Aloss_A5index}
\end{align}
where a flattened index equal to zero denotes the mode $(n,\ell,m) = (0,0,0)$.
Similarly, the gain and loss terms defined in Eqs.~\eqref{eq:C_Gain} and~\eqref{eq:C_Loss} are related to a six-index tensor, defined as
\begin{align}\label{eq:C_6index_original}
\begin{split}
    \tilde{C}_{ijksqw}
=\;&
    \int_{p_{1\mu},p_{2\mu}}
    \int_{\bp_{3},\bp_{4}}
    \tilde{\basisleft}_{i}^*(p_{1\mu})
    \tilde{\basisleft}_{j}^*(p_{2\mu})
    \frac{1}{p_1^0 p_2^0}
\\&\times
    \mW_{(\bp_1,\bp_2 \rightarrow \bp_3,\bp_4)}
    \, e^{\mathrm{s}_\mathrm{Q} \frac{P^\mu u_\mu}{\Lambda}}
\\&\times
    \tilde{\basis}_k(p_{1\mu})
    \tilde{\basis}_s(p_{2\mu})
    \tilde{\basis}_q(p_{3\mu})
    \tilde{\basis}_w(p_{4\mu})
    \,,
\end{split}
\end{align}
through the identities
\begin{align}
    C^{\text{gain}}_{ijks} =\;& 4\pi\, 
    \mathcal{Y}_{ii'}^* \mathcal{Y}_{jj'}^* \mathcal{Y}_{kk'} \mathcal{Y}_{ss'} 
    \tilde{C}_{ij00ks} \,, \label{eq:Cgain_C6index}\\
    C^{\text{loss}}_{ijks} =\;& 4\pi\, 
    \mathcal{Y}_{ii'}^* \mathcal{Y}_{jj'}^* \mathcal{Y}_{kk'} \mathcal{Y}_{ss'} 
    \tilde{C}_{i'j'k's'00} \,. \label{eq:Closs_C6index}
\end{align}
Beyond simplifying the notation, this unified formulation is also structurally compatible with the quantum generalization of the BBGKY hierarchy, that is, the inclusion of statistical factors accounting for Bose stimulation or Pauli blocking effects. To achieve this, we only need to redefine the statistical factor as $\mathrm{s}_\mathrm{Q} = 0$. Thus, these tensors not only streamline the numerical evaluation of collision integrals, but also provide a consistent foundation for future theoretical extensions.

We further note that a straightforward calculation yields a direct relationship between $\tilde{C}_{ijksqw}$ and $\tilde{A}_{iksqw}$ for massless particles, namely,
\begin{align}\label{eq:AC_Relationship}
\begin{split}
    \tilde{A}_{i ksqw}
    =
    \frac{\Lambda^3}{2\pi^{\frac{5}{2}}}
    \tilde{C}_{i0  ksqw} 
    \,.
\end{split}
\end{align}
Therefore, we demonstrate the symmetry-constrained relations by focusing on the structure of $\tilde{C}_{ijksqw}$. The symmetry-constrained relations for $\tilde{A}_{ijksq}$ in the massless case can be obtained directly by considering Eq.~\eqref{eq:AC_Relationship}. Although Eq.~\eqref{eq:AC_Relationship} no longer holds in the presence of mass, symmetry-constrained relations for $\tilde{A}_{ijksq}$ can still be derived in an analogous manner.

\vspace{3mm}
First, the integral~\eqref{eq:C_6index_original} should be invariant under a \textit{parity transformation} ($\boldsymbol{p} \to -\boldsymbol{p}$) applied to all momenta. Noting that $\mathrm{P}\, Y_\ell^m(\theta,\varphi) \,\mathrm{P} = (-1)^\ell Y_\ell^m(\theta,\varphi)$, it is straightforward to find
\begin{align}
\tilde{C}_{ijksqw} = (-1)^{\ell_i + \ell_j + \ell_k + \ell_s + \ell_q + \ell_w} \tilde{C}_{ijksqw} \,,
\end{align}
where $\ell_i$ is the polar index corresponding to the flattened index $i$, and likewise for the other $\ell$'s. Therefore, we obtain the first constraint,
\begin{align}
\tilde{C}_{ijksqw}=0, \;\;\text{if\;} \ell_i + \ell_j + \ell_k + \ell_s + \ell_q + \ell_w \mathrm{\; is \; odd} \,.
\end{align}

Second, Eq.~\eqref{eq:C_6index_original} should be invariant under a \textit{redefinition of the solid angle corresponding to an $O(3)$ rotation}, under which the complex spherical harmonics transform as
$
Y_{\ell}^m(\theta', \varphi') = \sum_{m^{\prime}=-\ell}^{\ell} \left[D_{m m^{\prime}}^{(\ell)}(\mathcal{R})\right]^* Y_{\ell}^{m^{\prime}}(\theta, \varphi)
$. Here, $D_{m m'}^{(\ell)}(\mathcal{R})$ is the Wigner D-matrix corresponding to the rotation $\mathcal{R}: (\theta, \varphi)\to(\theta', \varphi')$. The invariance relation reads
\begin{align}\label{eq:reduction_rotation}
\tilde{C}_{ijksqw} = \mD_{ijksqw,\, i'j'k's'q'w'} \, \tilde{C}_{i'j'k's'q'w'} \,,
\end{align}
where $\mD_{ijksqw,\, i'j'k's'q'w'}$ is the Kronecker product of six Wigner D-matrices, acting on each of the spherical harmonic indices,
\begin{align}
\begin{split}
&
    \mD_{ijksqw,\, i'j'k's'q'w'} 
=\;
    \left[D_{m_i m_{i'}}^{(\ell_i)}\right]
    \left[D_{m_j m_{j'}}^{(\ell_j)}\right]
\\\times\;&
    \left[D_{m_k m_{k'}}^{(\ell_k)}\right]^*
    \left[D_{m_s m_{s'}}^{(\ell_s)}\right]^*
    \left[D_{m_q m_{q'}}^{(\ell_q)}\right]^*
    \left[D_{m_w m_{w'}}^{(\ell_w)}\right]^* \,.
\end{split}
\end{align}
$m_i$-$m_w$ are the azimuthal indices corresponding to the flattened indices $i$-$w$, respectively.
The condition~\eqref{eq:reduction_rotation} requires that the integral tensor $\tilde{C}_{ijksqw}$ must lie within the null space of the operator $(\mD - I)$.

For a simple and explicit example, one may consider a special case of an $O(3)$ rotation---the \textit{redefinition of the azimuthal angle} ($\varphi \to \varphi + \Delta \varphi$). This angular translation leads to a phase shift in the complex spherical harmonics, $Y_{\ell}^m(\theta, \varphi + \Delta\varphi) = e^{\mathrm{i} m \Delta\varphi} Y_{\ell}^m(\theta, \varphi)$, as well as in the integral
\begin{align}
\begin{split}
&
    \tilde{C}_{ijksqw} 
    e^{\mathrm{i}(-m_i - m_j + m_k + m_s + m_q + m_w)\Delta\varphi} \tilde{C}_{ijksqw} \,.
\end{split}
\end{align}
Such rotational invariance yields the constraint
\begin{align}
\tilde{C}_{ijksqw}=0, \;\; \text{if\;} m_i + m_j \neq  m_k + m_s + m_q + m_w  \,.
\label{eq:O2_constrain}
\end{align}

There is another simple constraint that can be regarded as a subset of the $O(3)$ rotational symmetry conditions. That is, the polar angle indices $\ell_i, \ell_j, \ell_k, \ell_s, \ell_q, \ell_w$ in $\tilde{C}_{ijksqw}$ must satisfy a generalized polygon inequality. Here, ``generalized'' means that one or more side lengths may vanish. The polygon inequality requires that any one of the six angular momenta must be less than the sum of the remaining five,
\begin{align}\label{eq:generalized_polygon_inequality}
\begin{split}
    \ell_i + \ell_j + \ell_k + \ell_s + \ell_q + \ell_w 
    > 2\max(\ell_i,\ell_j, \ell_k, \ell_s, \ell_q, \ell_w)\,.
\end{split}
\end{align}
This constraint resembles the coupling of angular momenta, due to the fact that $\tilde{C}_{ijksqw}$ shares mathematically equivalent properties with the Clebsch--Gordan coefficients.

In general, the other constraint equations arising from Eq.~\eqref{eq:reduction_rotation} are not as simple as Eqs.~\eqref{eq:O2_constrain} or~\eqref{eq:generalized_polygon_inequality}. 
We take a general Euler angle in $\mathcal{R}$ and solve the eigenvalue problem of $\mD$. If the matrix $(\mD - I)$ admits $s$ independent null-space vectors $w_1, w_2, \dots, w_s$, each of dimension $r$, then it suffices to compute the integral factors for only $s$ carefully chosen parameter sets, from which all others can be reconstructed.
We define such a set of specific parameters as follows.
Let the $s$ null-space vectors $\{w_i\}$ form an $r \times s$ matrix $W = (w_1, w_2, \dots, w_s)$. 
If the submatrix composed of rows $r_1, r_2, \dots, r_s$ of $W$ is full rank, then the parameters corresponding to those selected rows constitute a valid specific parameter set. 

\vspace{3mm}
These symmetry constraints---parity and rotational invariance---not only ensure consistency with fundamental physical principles, but also dramatically reduce the number of independent integral factors that must be computed. Together, they provide a systematic and efficient framework for minimizing computational cost while preserving the full structure and accuracy of the collision terms.

\subsection{Integral Form of the Collision Kernel}\label{sec:Collision Kernel:Integral Form of the Collision Kernel}

In principle, the high-dimensional integrals $\tilde{A}_{ijksq}$ and $\tilde{C}_{ijksqw}$, i.e., Eqs.~\eqref{eq:A_5index_original} and~\eqref{eq:C_6index_original}, that appear in the spectral BBGKY hierarchy can be evaluated numerically.
However, we find that the evolution equations are highly sensitive to the precision of the collision integrals $A$ and $C$. These are eight-fold integrals and are computationally expensive.
To improve both the accuracy and efficiency of these integrals, we develop a reduction scheme that transforms the original eight-dimensional expressions into three-dimensional ones, which will be introduced in this subsection.
Under grid-based integration schemes, the number of function evaluations is reduced from $H^8$ to $H^3$, where $H$ is the number of sampling points per dimension.
For instance, with a typical resolution of $H = 100$, the total number of evaluations drops from $10^{16}$ to $10^6$, yielding a computational speedup of ten orders of magnitude.
This reduction also results in a substantial improvement in numerical accuracy. Given a fixed computational budget of $10^6$ function evaluations, the original eight-dimensional formulation permits a grid resolution of only $H \approx 5.6$ per dimension, whereas the reduced three-dimensional form supports $H = 100$.
This significantly facilitates analytic evaluation in the massless case (Sec.~\ref{subsec:Summation Form of the Collision Kernel}) and enhances numerical precision in the massive case.

The procedure for reducing the integral from eight dimensions to three is presented below.

Since the distribution functions are Lorentz scalars, the values of integrals constructed from these functions remain unchanged across reference frames related by Lorentz boosts. To take advantage of this property, we first rewrite the spherical harmonics as polynomials using Eqs.~\eqref{eq:ReducibleTesnor_begin}-\eqref{eq:ReducibleTesnor_end}.
It is useful to express this in a compact manner,
\begin{align}
\begin{split}
    |\bp|^\ell Y_\ell^m(\theta, \phi) 
=\;&
    \sum_{\rho=0}^{\lfloor \frac{\ell - |m|}{2} \rfloor} 
    \sum_{\tau=0}^{|m|} 
    \sum_{\eta=0}^{\rho}
    \sum_{\epsilon=0}^{\rho-\eta}
    \mI_{\alpha} (\ell,m,\rho,\tau,\eta,\epsilon)
\\&\times
    p_x^{\tau+2\eta} 
    p_y^{|m|-\tau+2\epsilon}
    p_z^{\ell - |m| - 2\tau - 2\epsilon}
    \,,
\end{split}
\end{align}
where
\begin{align}
\begin{split}
&
    \mI_{\alpha} (\ell,m,\rho,\tau,\eta,\epsilon)
\\=\;&
    2^{-\ell} \sqrt{\frac{2\ell+1}{4\pi}}\, 
    \sqrt{\frac{(\ell - |m|)!}{(\ell + |m|)!}} 
    \frac{(-1)^{\rho+|m|}}{(\ell-\rho)!} 
\\&\times
    \frac{(2\ell - 2\rho)!}{(\ell - 2\rho - |m|)!}
    \frac{|m|! \; \mathrm{i}^{|m|-\tau}}{\tau!(|m|-\tau)!}
    \frac{(-1)^{\tau\,\vartheta(-\frac{1}{2}-m)}}{\eta!\epsilon!(\rho-\eta-\epsilon)!}
    \,,
\end{split}
\end{align}
with $\vartheta(x)$ denoting the Heaviside step function, and $\lfloor x \rfloor$ denoting the greatest integer not greater than $x$.

Next, we can transform the integral $\tilde{C}_{ijksqw}$ into the center-of-momentum (CM) frame.
We define a Lorentz transformation matrix $\mathcal{M}$, which maps four-vectors from the frame $S$ to another frame that satisfies two conditions: it is comoving with the colliding particle pair, and the time-like vector $u^\mu$ is aligned with the spatial $z$-axis in this frame.
This transformation consists of a boost corresponding to the former, followed by a spatial rotation for the latter,
\begin{align}
\begin{split}
    {\mathcal{M}^{\alpha}}_{\beta}
=\;&
    {{\Lambda_\mathrm{boost}}^{\alpha}}_{\xi}\;
    {{\Lambda_\mathrm{rotation}}^{\xi}}_{\beta}\,,
\end{split}
\end{align}
with
\begin{align}
\begin{split}
&
    \big({{\Lambda_\mathrm{boost}}^{\alpha}}_{\xi}\big)
\\=\;&
    \frac{1}{P}
    \begin{pmatrix}
    P^0
    &\,\,\,\, -\hat{P}^i|\bP|
    \\
    -\hat{P}^i|\bP|
    & \,\,\,\,
    P\delta_{ij}+(P^0-P)\hat{P}^i \hat{P}^j
    \end{pmatrix}\,,
\end{split}
\\
\begin{split}
&
    \big({{\Lambda_\mathrm{rotation}}^{\xi}}_{\beta}\big)
\\=\;&
    \left(\begin{array}{cccc}
    1 & 0 & 0 & 0 \\
    0 & \cos \theta_P \cos \phi_P & -\sin \phi_P & \sin \theta_P \cos \phi_P \\
    0 & \cos \theta_P \sin \phi_P & \cos \phi_P & \sin \theta_P \sin \phi_P \\
    0 & -\sin \theta_P & 0 & \cos \theta_P
    \end{array}\right)\,.
\end{split}
\end{align}
Here, $P^\mu$ is the (conserved) total four-momentum of the incoming and outgoing particles, defined as $P^\mu = p_1^\mu + p_2^\mu = p_{3}^{\mu} + p_{4}^{\mu}$. $P = \sqrt{P^\mu P_\mu}$, and $\hat{\bP}$ denotes the unit vector in the direction of the spatial part of $P^\mu$ in the frame $S$, with $\theta_P$ and $\phi_P$ denoting the polar and azimuthal angles of $\hat{\bP}$, respectively.
The integral $\tilde{C}_{ijksqw}$ then becomes
\begin{widetext}
\begin{align}
\begin{split}
&
    \tilde{C}_{ijksqw}
\\=\;&
    \int_{\bp_{1},\bp_{2},\bp_{3},\bp_{4}}
    \frac{(2\pi)^6 n_i!\,n_j!\,\mW_{(\bp_1,\bp_2 \rightarrow \bp_3,\bp_4)}e^{(\mathrm{s}_\mathrm{Q}-2) \frac{P^\mu u_\mu}{\Lambda}}}{\Lambda^4\,(2\ell_i+n_i+2)!\,(2\ell_j+n_j+2)!}  
    \left(\frac{u\cdot p_{1}}{\Lambda\,|\bp_1|} \right)^{\ell_i+\ell_k+1}  
    \left( \frac{u\cdot p_{2}}{\Lambda\,|\bp_2|} \right)^{\ell_j+\ell_s+1}  
    \left( \frac{u\cdot p_{3}}{\Lambda\,|\bp_3|} \right)^{\ell_q} 
    \left( \frac{u\cdot p_{4}}{\Lambda\,|\bp_4|} \right)^{\ell_w} 
\\\times\;&
    L_{n_i}^{(2\ell_i+2)}\left( \frac{u\cdot p_{1}}{\Lambda} \right)\;\; 
    L_{n_j}^{(2\ell_j+2)}\left( \frac{u\cdot p_{2}}{\Lambda} \right)\;\;  
    L_{n_k}^{(2\ell_k+2)}\left( \frac{u\cdot p_{1}}{\Lambda} \right)\;\;  
    L_{n_s}^{(2\ell_s+2)}\left( \frac{u\cdot p_{2}}{\Lambda} \right)\;\;  
    L_{n_q}^{(2\ell_q+2)}\left( \frac{u\cdot p_{3}}{\Lambda} \right)\;\;   
    L_{n_w}^{(2\ell_w+2)}\left( \frac{u\cdot p_{4}}{\Lambda} \right) 
\\\times\;&
    \bigg[
    \hatprod\limits_{\chi=i,j}
    \sum_{\rho_{\chi}=0}^{\lfloor \frac{\ell_{\chi} - |m_{\chi}|}{2} \rfloor} 
    \sum_{\tau_{\chi}=0}^{|m_{\chi}|} 
    \sum_{\eta_{\chi}=0}^{\rho_{\chi}}
    \sum_{\epsilon_{\chi}=0}^{\rho_{\chi}-\eta_{\chi}}
    \mI^*_{\alpha} (\ell_{\chi},m_{\chi},\rho_{\chi},\tau_{\chi},\eta_{\chi},\epsilon_{\chi})
    \bigg]
    \bigg[
    \hatprod_{\chi=k,s,q,w}
    \sum_{\rho_{\chi}=0}^{\lfloor \frac{\ell_{\chi} - |m_{\chi}|}{2} \rfloor} 
    \sum_{\tau_{\chi}=0}^{|m_{\chi}|} 
    \sum_{\eta_{\chi}=0}^{\rho_{\chi}}
    \sum_{\epsilon_{\chi}=0}^{\rho_{\chi}-\eta_{\chi}}
    \mI_{\alpha} (\ell_{\chi},m_{\chi},\rho_{\chi},\tau_{\chi},\eta_{\chi},\epsilon_{\chi})
    \bigg]
\\\times\;&
    \bigg[
    \hatprod_{\chi = i,j,k,s,q,w}
    \bigg(\prod_{\beta_{\chi}=1}^{\tau_{\chi}+2\eta_{\chi}}\mathcal{M}_{1 \mu_{\beta_{\chi}}}\bigg)
    \bigg(\prod_{\beta_{\chi}=\tau_{\chi}+2\eta_{\chi} + 1}^{|m_{\chi}|+2\epsilon_{\chi}+2\eta_{\chi}}\mathcal{M}_{2 \mu_{\beta_{\chi}}}\bigg)
    \bigg(\prod_{\beta_{\chi}=|m_{\chi}|+2\epsilon_{\chi}+2\eta_{\chi} + 1}^{\ell_{\chi}}\mathcal{M}_{3 \mu_{\beta_{\chi}}}\bigg)
    \bigg]
    p_{1}^{\mu_{\beta_{i}}} \;
    p_{2}^{\mu_{\beta_{j}}} \;
    p_{1}^{\mu_{\beta_{k}}}\;
    p_{2}^{\mu_{\beta_{s}}} \;
    p_{3}^{\mu_{\beta_{q}}}\;
    p_{4}^{\mu_{\beta_{w}}} 
    \,,
\end{split}\label{eq:step2_C}
\end{align}
\end{widetext}
where $a \cdot b \equiv g^{\mu\nu} a_\mu b_\nu$, with the metric convention $g^{\mu\nu} = \mathrm{diag}(+1,-1,-1,-1)$, and $\hatprod$ is defined in Eq.~\eqref{eq:hatprod}.
The reformulation in Eq.~\eqref{eq:step2_C} allows us to extend the techniques developed by de Groot~\cite{DeGroot:1980dk}, pp.~373-380, in order to reduce the dimensionality of the integral.

We introduce the following change of variables,
\begin{gather}
P^\mu = p_1^\mu + p_2^\mu = p_3^\mu + p_4^\mu = P^{\prime \mu} \,, \label{eq:relationship_P_Pprime}\\
Q^\mu = \Delta_P^{\mu \nu} (p_{1 \nu} - p_{2 \nu})\,, 
\\
Q^{\prime \mu} = \Delta_P^{\mu \nu} (p_{3 \nu} - p_{4 \nu})\,,
\end{gather}
where $\Delta_P^{\mu \nu} = g^{\mu \nu} - \frac{P^\mu P^\nu}{P^2}$ is the transverse projector with respect to the total four-momentum $P^\mu$.
The four-momenta of the individual particles can then be reconstructed as
\begin{align}
\begin{split}
    p_1^\mu &= \frac{P^\mu +  Q^\mu}{2}  \,, \qquad
    p_2^\mu = \frac{P^\mu -  Q^\mu}{2}  \,, \\
    p_3^\mu &= \frac{P^\mu +  Q'^\mu}{2}  \,, \qquad
    p_4^\mu = \frac{P^\mu -  Q'^\mu}{2}  \,.
\end{split}
\end{align}
Using the orthogonality conditions and the simplified expressions for the Lorentz scalars $Q^2 \equiv Q^\mu Q_\mu$ and $Q'^2 \equiv Q'^\mu Q'_\mu$,
\begin{align}
&   P^\mu Q_\mu=0, \qquad P^\mu Q_\mu^{\prime}=0 \,,\\
&    Q^2 = Q'^2 = - P^2 + 4\mass^2 \,,
\end{align}
we can now determine the integration element in terms of the new variables,
\begin{align}
    \mathrm{d} \mu(P)
=\;&
    \mathrm{d}^4 P \theta\left(P^0\right) \theta\left(P^2-4\mass^2\right)\,, 
\\
    \mathrm{d} \mu(P')
=\;&
    \mathrm{d}^4 P' \theta\left(P'^0\right) \theta\left(P'^2-4\mass^2\right)\,, 
\\
    \mathrm{d} \mu(Q)
=\;&
    \mathrm{d}^4 Q \delta(Q \cdot P) 
    \delta\left(Q^2+ P^2-4\mass^4\right)\,,
\\
    \mathrm{d} \mu(Q')
=\;&
    \mathrm{d}^4 Q' \delta(Q' \cdot P) 
    \delta\left(Q'^2+ P^2-4\mass^2\right)\,.
\end{align}
The integration elements can be further simplified to
\begin{align} 
d \mu(Q) & =
\sin\theta d\theta d\phi
\frac{\sqrt{-Q^2}}{2 P}\,, \label{eq:simplify_muQ}
\\
d \mu\left(Q^{\prime}\right) & = 
\sin\theta'd\theta' d\phi'
\frac{\sqrt{-Q^{\prime 2}}}{2 P}
\label{eq:simplify_muQprime}
\,,
\end{align}
where $(\theta, \phi)$ and $(\theta', \phi')$ denote the polar and azimuthal angles of $\bQ$ and $\bQ'$, respectively.
The Jacobian determinant is
\begin{align}
\begin{split}
&
\frac{\mathrm{d}^3 p_1}{p_1^0} 
\frac{\mathrm{d}^3 p_2}{p_2^0} 
\frac{\mathrm{d}^3 p_3}{p_3^0} 
\frac{\mathrm{d}^3 p_4}{p_4^0} 
=
\mathrm{d} \mu(P) 
\mathrm{d} \mu(P')
\mathrm{d} \mu(Q) 
\mathrm{d} \mu(Q')
\,.
\end{split}
\end{align}
The transition rate can also be written as a function of the total and relative four-momenta,
\begin{align}
\begin{split}
    &W_{(\bp_1,\bp_2 \rightarrow \bp_{3},\bp_{4})}
=
    P^2 \sigma\left(P,\cos \Theta\right)
    \,,
\end{split}
\end{align}
where the differential cross-section $\sigma(P, \cos \Theta)$ depends on the magnitude $P$ of the total four-momentum and the angle $\Theta$ between the incoming and outgoing relative momenta. The angle is defined by
\begin{align}
\cos \Theta=\hat{\boldsymbol{Q}} \cdot \hat{\boldsymbol{Q}}^{\prime}\,,
\end{align}
where 
\begin{align}
    \hat{Q}^\mu=Q^\mu / \sqrt{-Q^2}
    \,.
\end{align}

Finally, to simplify the angular dependence, the cross-section is expanded into Legendre polynomials,
\begin{align}\label{eq:sigma_legendre_expansion}
\begin{split}
&
    \sigma(P, \cos \Theta) 
=
    \frac{1}{\Lambda^2} \sum_{g=0}^{\infty} \sigma^{(g)}\left(\frac{P}{\Lambda}\right) P_g(\cos \Theta)
    \,,
\end{split}
\end{align}
where the dimensionless expansion coefficients $\sigma^{(g)}$ are defined as
\begin{align}
\begin{split}
&
    \sigma^{(g)}\left(\frac{P}{\Lambda}\right) 
=
    \frac{2g + 1}{2} \Lambda^2
     \int_{-1}^1 \mathrm{d}x\, P_g(x)\, \sigma(P, x) \,.
\end{split}
\end{align}
Furthermore, by applying the addition theorem of spherical harmonics, the angular structure can be rewritten as
\begin{align}
    P_g(\cos \Theta) 
= 
    \frac{4\pi}{2g + 1} \sum_{h=-g}^{g} \left[Y_g^h(\theta, \phi)\right]^* Y_g^h(\theta', \phi') 
    \,.
\end{align}
Here, $(\theta, \phi)$ and $(\theta', \phi')$ denote the polar and azimuthal angles of $\bQ$ and $\bQ'$, respectively, as introduced earlier.
The integral $\tilde{C}_{ijksqw}$ can then be finally expressed as a three-fold integral,
\begin{widetext}
\begin{align}
\begin{split}
&
    \tilde{C}_{ijksqw}
\\=\;&
    \bigg[
    \hatprod_{\chi=i,j}
    \sum_{\rho_{\chi}=0}^{\lfloor \frac{\ell_{\chi} - |m_{\chi}|}{2} \rfloor} 
    \sum_{\tau_{\chi}=0}^{|m_{\chi}|} 
    \sum_{\eta_{\chi}=0}^{\rho_{\chi}}
    \sum_{\epsilon_{\chi}=0}^{\rho_{\chi}-\eta_{\chi}}
    \mI^*_{\alpha} (\ell_{\chi},m_{\chi},\rho_{\chi},\tau_{\chi},\eta_{\chi},\epsilon_{\chi})
    \bigg]
    \bigg[
    \hatprod_{\chi=k,s,q,w}
    \sum_{\rho_{\chi}=0}^{\lfloor \frac{\ell_{\chi} - |m_{\chi}|}{2} \rfloor} 
    \sum_{\tau_{\chi}=0}^{|m_{\chi}|} 
    \sum_{\eta_{\chi}=0}^{\rho_{\chi}}
    \sum_{\epsilon_{\chi}=0}^{\rho_{\chi}-\eta_{\chi}}
    \mI_{\alpha} (\ell_{\chi},m_{\chi},\rho_{\chi},\tau_{\chi},\eta_{\chi},\epsilon_{\chi})
    \bigg]
\\\times&
    \frac{1}{(2\pi)^2\Lambda^6} 
    \frac{n_i!}{(2\ell_i+n_i+2)!}  
    \frac{n_j!}{(2\ell_j+n_j+2)!} 
    \int_0^\infty d|\bP| |\bP|^2 
    \int_{|\bP|}^{\infty} d P^0 
    \frac{\sqrt{-Q^2}}{2} 
    \frac{\sqrt{-Q'^2}}{2}
    e^{(\mathrm{s}_\mathrm{Q}-2)\frac{P_{\mu} u^\mu}{\Lambda}} 
     \sum_{g=0}^{\infty} \sigma^{(g)}\left(\frac{P}{\Lambda}\right)     \frac{4\pi}{2g + 1} \sum_{h=-g}^{g} 
\\\times&
    \int_0^{2\pi} d\phi_P \int_0^\pi d\theta_P \sin\theta_P
    \bigg[
    \hatprod_{\chi = i,j,k,s,q,w}
    \bigg(\prod_{\beta_{\chi}=1}^{\tau_{\chi}+2\eta_{\chi}}\mathcal{M}_{1 \mu_{\beta_{\chi}}}\bigg)
    \bigg(\prod_{\beta_{\chi}=\tau_{\chi}+2\eta_{\chi} + 1}^{|m_{\chi}|+2\epsilon_{\chi}+2\eta_{\chi}}\mathcal{M}_{2 \mu_{\beta_{\chi}}}\bigg)
    \bigg(\prod_{\beta_{\chi}=|m_{\chi}|+2\epsilon_{\chi}+2\eta_{\chi} + 1}^{\ell_{\chi}}\mathcal{M}_{3 \mu_{\beta_{\chi}}}\bigg)
    \bigg]
\\\times&
    \bigg[\int_0^\pi \sin \theta d \theta
    \left( \frac{u\cdot p_{1}}{\Lambda\,|\bp|_1} \right)^{\ell_i+\ell_k+1}  
    \left( \frac{u\cdot p_{2}}{\Lambda\,|\bp|_2} \right)^{\ell_j+\ell_s+1}  
    L_{n_i}^{(2\ell_i+2)}\left( \frac{u\cdot p_{1}}{\Lambda} \right) 
    L_{n_j}^{(2\ell_j+2)}\left( \frac{u\cdot p_{2}}{\Lambda} \right) 
\\\times&
    L_{n_k}^{(2\ell_k+2)}\left( \frac{u\cdot p_{1}}{\Lambda} \right) 
    L_{n_s}^{(2\ell_s+2)}\left( \frac{u\cdot p_{2}}{\Lambda} \right)
    \int_0^{2\pi} d \phi
    p_{1}^{\mu_{\beta_{i}}} 
    p_{2}^{\mu_{\beta_{j}}} 
    p_{1}^{\mu_{\beta_{k}}} 
    p_{2}^{\mu_{\beta_{s}}} 
     \left[Y_g^h(\theta, \phi)\right]^* \bigg]
\\\times&
    \bigg[\int_0^\pi \sin \theta' d \theta'
    \left( \frac{u\cdot p_{3}}{\Lambda\,|\bp|_3} \right)^{\ell_q}  
    \left( \frac{u\cdot p_{4}}{\Lambda\,|\bp|_4} \right)^{\ell_w}  
    L_{n_q}^{(2\ell_q+2)}\left( \frac{u\cdot p_{3}}{\Lambda} \right) 
    L_{n_w}^{(2\ell_w+2)}\left( \frac{u\cdot p_{4}}{\Lambda} \right) 
    \int_0^{2\pi} d \phi'
    p_{3}^{\mu_{\beta_{q}}}
    p_{4}^{\mu_{\beta_{w}}} 
     Y_g^h(\theta', \phi') \bigg]
    \,,
\end{split}\label{eq:step5_C}
\end{align}
where $(\theta_P, \phi_P)$, $(\theta, \phi)$, and $(\theta', \phi')$ denote the polar and azimuthal angles of the spatial parts of $P^\mu$, $Q^\mu$, and $Q'^\mu$, respectively.
The integral $A$ in three dimensions is
\begin{align}
\begin{split}
&
    \tilde{A}_{ijksq}
\\=\;&
    \bigg[
    \hatprod_{\chi=i}
    \sum_{\rho_{\chi}=0}^{\lfloor \frac{\ell_{\chi} - |m_{\chi}|}{2} \rfloor} 
    \sum_{\tau_{\chi}=0}^{|m_{\chi}|} 
    \sum_{\eta_{\chi}=0}^{\rho_{\chi}}
    \sum_{\epsilon_{\chi}=0}^{\rho_{\chi}-\eta_{\chi}}
    \mI^*_{\alpha} (\ell_{\chi},m_{\chi},\rho_{\chi},\tau_{\chi},\eta_{\chi},\epsilon_{\chi})
    \bigg]
    \bigg[
    \hatprod_{\chi=j,k,s,q}
    \sum_{\rho_{\chi}=0}^{\lfloor \frac{\ell_{\chi} - |m_{\chi}|}{2} \rfloor} 
    \sum_{\tau_{\chi}=0}^{|m_{\chi}|} 
    \sum_{\eta_{\chi}=0}^{\rho_{\chi}}
    \sum_{\epsilon_{\chi}=0}^{\rho_{\chi}-\eta_{\chi}}
    \mI_{\alpha} (\ell_{\chi},m_{\chi},\rho_{\chi},\tau_{\chi},\eta_{\chi},\epsilon_{\chi})
    \bigg]
\\\times&
    \frac{1}{(2\pi)^5 \Lambda^4}
    \frac{n_i!}{(2\ell_i+n_i+2)!}  
    \int_0^\infty d|\bP| |\bP|^2 
    \int_{|\bP|}^{\infty} d P^0 
    \frac{\sqrt{-Q^2}}{2} 
    \frac{\sqrt{-Q'^2}}{2}
    e^{(\mathrm{s}_\mathrm{Q} -2) \frac{P^\mu u_\mu}{\Lambda}}
    \sum_{g=0}^{\infty} \sigma^{(g)}\left(\frac{P}{\Lambda}\right)     \frac{4\pi}{2g + 1} \sum_{h=-g}^{g}
\\\times&
    \int_0^{2\pi} d\phi_P \int_0^\pi d\theta_P \sin\theta_P
    \bigg[
    \hatprod_{\chi = i,j,k,s,q}
    \bigg(\prod_{\beta_{\chi}=1}^{\tau_{\chi}+2\eta_{\chi}}\mathcal{M}_{1 \mu_{\beta_{\chi}}}\bigg)
    \bigg(\prod_{\beta_{\chi}=\tau_{\chi}+2\eta_{\chi} + 1}^{|m_{\chi}|+2\epsilon_{\chi}+2\eta_{\chi}}\mathcal{M}_{2 \mu_{\beta_{\chi}}}\bigg)
    \bigg(\prod_{\beta_{\chi}=|m_{\chi}|+2\epsilon_{\chi}+2\eta_{\chi} + 1}^{\ell_{\chi}}\mathcal{M}_{3 \mu_{\beta_{\chi}}}\bigg)
    \bigg]
\\\times&
    \bigg[
    \int_0^\pi \sin \theta d \theta
    \left( \frac{u\cdot p_{1}}{\Lambda|\bp_1|}\right)^{\ell_i+\ell_j+1}  
    \left( \frac{u\cdot p_{2}}{\Lambda|\bp_2|} \right)^{\ell_k} 
    L_{n_i}^{(2\ell_i+2)}\left( \frac{u\cdot p_{1}}{\Lambda} \right) 
    L_{n_j}^{(2\ell_j+2)}\left( \frac{u\cdot p_{1}}{\Lambda} \right) 
    L_{n_k}^{(2\ell_k+2)}\left( \frac{u\cdot p_{2}}{\Lambda} \right) 
\\\times&
    \int_0^{2\pi} d \phi
    p_{1}^{\mu_{\beta_{i}}} 
    p_{1}^{\mu_{\beta_{j}}}
    p_{2}^{\mu_{\beta_{k}}} 
    \left[Y_g^h(\theta, \phi)\right]^*
    \bigg]
\\\times&
    \bigg[
    \int_0^\pi \sin \theta' d \theta'
    \left( \frac{u\cdot p_{3}}{\Lambda|\bp_3|} \right)^{\ell_s}  
    \left( \frac{u\cdot p_{4}}{\Lambda|\bp_4|} \right)^{\ell_q}  
    L_{n_s}^{(2\ell_s+2)}\left( \frac{u\cdot p_{3}}{\Lambda} \right) 
    L_{n_q}^{(2\ell_q+2)}\left( \frac{u\cdot p_{4}}{\Lambda} \right) 
    \int_0^{2\pi} d \phi
    p_{3}^{\mu_{\beta_{s}}}
    p_{4}^{\mu_{\beta_{q}}} 
    Y_g^h(\theta', \phi') 
    \bigg]
    \,,
\end{split}\label{eq:step5_A}
\end{align}
\end{widetext}

The resulting three-dimensional formulation in Eqs.~\eqref{eq:step5_C}-\eqref{eq:step5_A} retains full generality, making it applicable to both massive and massless particles, as well as to arbitrary collision kernels.
It is therefore useful for both analytical and numerical studies in kinetic theory.

The analytic expression for the free-streaming integral $\boldsymbol{B}_{ij}$, given in Eq.~\eqref{eq:Free streaming Integral}, can be readily evaluated.
For convenience, we define
\begin{align}
\begin{split}
    B_{ij,m_0}
=\;&
    \int_{p_\mu} 
    \basisleft_{i}(p_{\mu})
    \basis_{j}(p_{\mu})
    Y_{1,m_0}(\theta,\phi)
    \;,
\end{split}
\end{align}
where $m_0 = +1$, $-1$, and $0$ correspond to the $x$, $y$, and $z$ components of the vector $\boldsymbol{B}$, respectively.

The real-valued representation of $B_{ij,m_0}$ is related to its complex-valued representation $\tilde{B}_{ij,m_0}$ via the transformation
\begin{align}
    B_{ij,m_0} =\;&\mathcal{Y}_{ii'}^* \mathcal{Y}_{jj'} \mathcal{Y}_{m_0 m'_0} \tilde{B}_{i'j',m'_0} \,,
\end{align}
where $Y_\ell^{m_0} = \mathcal{Y}_{m_0 m'_0} Y_{\ell,m'_0}$, and the matrix $\mathcal{Y}_{m_0 m'_0}$ corresponds to the transformation between real and complex spherical harmonics for $\ell = 1$ (see Appendix~\ref{Appendix:Spherical Harmonics} Eq.~\eqref{eq:SH_real_vs_complex}).
The complex-valued representation $\tilde{B}_{ij,m_0}$ can be calculated by
\begin{align}
\begin{split}
    \tilde{B}_{ij,m_0}
=\;&
    \frac{n_i!}{(2\ell_i+n_i+2)!} 
    \delta_{m_0+m_i+m_j,0}
    (
    \delta_{\ell_i,\ell_j-1}
    +
    \delta_{\ell_i,\ell_j+1}
    ) 
\\&\times
    \sum_{k_i=0}^{n_i}
    \frac{(-1)^{k_i}(n_i+2\ell_i+2)!}{(n_i-k_i)!(k_i+2\ell_i+2)! k_i!}
\\&\times
    \sum_{k_j=0}^{n_j}
    \frac{(-1)^{k_j} (n_j+2\ell_j+2)!}{(n_j-k_j)! (k_j+2\ell_j+2)! k_j!}
\\&\times
    (\ell_i+\ell_j+k_i+k_j+2)!
    \sqrt{\frac{3\left(2 \ell_i+1\right)\left(2 \ell_j+1\right)}{4 \pi}}
\\&\times
    \left(\begin{array}{ccc}
    1 & \ell_i & \ell_j \\
    0 & 0 & 0
    \end{array}\right)
    \left(\begin{array}{ccc}
    1 & \ell_i & \ell_j \\
    m_0 & m_i & m_j
    \end{array}\right)
    \,,
\end{split}
\end{align}
where the first two factors in the second line are $3j$ symbols; the rules for their evaluation are given in Appendix~\ref{Appendix:3j_symbol}.

In addition to the reduction in computational cost mentioned at the beginning of this subsection, the spectral decomposition in Eq.~\eqref{eq:sigma_legendre_expansion} enables a modular treatment of the collision integrals. Since the collision integrals $A$ and $C$ depend linearly on the collision kernel $\mathcal{W}$, i.e.,
\begin{align}
C[\mathcal{W}_1 + \mathcal{W}_2] = C[\mathcal{W}_1] + C[\mathcal{W}_2]\,,
\end{align}
the total contribution for any cross section can be constructed as a linear combination of the basis components $\sigma^{(g)}$. This allows precomputed building blocks to be reused.

The temperature dependence of the collision integrals $A$ and $C$ can be readily determined.
Using $[A]$ to denote the dimensionality of a physical quantity $A$, we have
\begin{align}
[A] = [\Lambda], \quad [C] = [\Lambda]^{-2} \,.
\end{align}
For a conformal system, where the temperature can always be chosen as the energy unit and all quantities with dimensions can be rescaled accordingly, this implies that correlations play a more significant role at low temperatures, while their impact becomes less pronounced at high temperatures.
Furthermore, once the integrals have been evaluated at a specific scale $\Lambda$, the results at arbitrary temperatures can be obtained through dimensional analysis, facilitating both the generalization and reuse of the computed integrals.

\subsection{Summation Form of the Collision Kernel}\label{subsec:Summation Form of the Collision Kernel}

Further simplifications can be made in addition to the dimensional reduction discussed in Sec.~\ref{sec:Collision Kernel:Integral Form of the Collision Kernel}. 
For a broad class of differential cross sections involving massless particles, the collision integrals can be evaluated exactly in closed analytic form, without the need for numerical integration.

This class includes cases where $\sigma^{(g)}(P/\Lambda)$ is either a polynomial in $P/\Lambda$ or can be well approximated by one. As an example, we present the final result for the case
\begin{align}\label{eq:example_collision_kernel}
W_{(p_1^\mu,p_2^\mu \to p_{3}^{\mu},p_{4}^{\mu})} = \frac{(p_1^\mu + p_2^\mu)^2}{\Lambda^2} \, \sigma^{(0)}(P/\Lambda),
\end{align}
where $\sigma^{(0)}(P/\Lambda)$ is taken to be a constant.

The key steps in the procedure are as follows. 
First, the Lorentz transformation matrix $\mathcal{M}$ is substituted. 
Second, the generalized Laguerre polynomials are expressed in closed form [Eq.~\eqref{eq:laguerrepolynomials}]. 
Third, the angular integrations over the relative momenta---specifically, $(\theta, \phi)$ and $(\theta', \phi')$---are performed. 
Fourth, the dependence on the total momentum $P^\mu$ is integrated out. 
Finally, the summation order is reduced to obtain a computationally tractable expression.

The final analytical expression for the integral factor 
$\tilde{C}$ is given by
\begin{widetext}
\begin{align}\label{eq:C_result}
\begin{split}
&
    \tilde{C}_{n_a l_a m_a, n_b l_b m_b, n_1 \ell_1 m_1, n_2 \ell_2 m_2, n_{3} \ell_{3} m_{3}, n_{4} \ell_{4} m_{4}}
\\=\;&
    \frac{\sigma^{(0)}}{\pi^2 \Lambda^2} 
    \bigg[\prod_{i\;\in\{a,b\}}
    \frac{n_i !}{(2l_i+n_i+2)!}\bigg]
    \bigg[\prod_{i\;\in\{a,b,1,2,3,4\}}
    \sqrt{2 \ell_i+1} 
    \left[\frac{(\ell_i-|m_i|)!}{(\ell_i+|m_i|)!}\right]^{1 / 2}  
    (-1)^{\frac{m_i+|m_i|}{2}}\bigg] 
    \sum_{w'_{1}=0}^{
    \lfloor\frac{\ell_{a}-|m_{a}|}{2}\rfloor
    +
    \lfloor\frac{\ell_{1}-|m_{1}|}{2}\rfloor}
    \sum_{p'_{1}=0}^{|m_{a}|+|m_{1}|}
\\&
    \sum_{w'_{2}=0}^{
    \lfloor\frac{\ell_{b}-|m_{b}|}{2}\rfloor
    +
    \lfloor\frac{\ell_{2}-|m_{2}|}{2}\rfloor}
    \sum_{p'_{2}=0}^{|m_{b}|+|m_{2}|}
    \sum_{f'_1=0}^{n_a+n_1}
    \sum_{f'_2=0}^{n_b+n_2}
    \Big[
    \hatprod_{i \;\in \{1,2\}}
    \sum_{\eta'_{i}=0}^{w'_{i}} 
    \sum_{\epsilon'_{i}=0}^{w'_{i}-\eta'_{i}}
    \Big]
    \sum_{t''_{12}=0}^{f'_1+f'_2}
    \Big[
    \hatprod_{i \;\in \{3,4\}}
    \sum_{w_{i}=0}^{\lfloor\frac{\ell_{i}-|m_{i}|}{2}\rfloor}
    \sum_{p_{i}=0}^{|m_{i}|}
    \sum_{\eta_{i}=0}^{w_{i}} 
    \sum_{\epsilon_{i}=0}^{w_{i}-\eta_{i}}
    \sum_{f_i=0}^{n_i}
    \Big]
    \sum_{t_{34}=0}^{f_3+f_4}
\\&
    \sum_{x''_{i,12},x''_{j,12},x''_{k,12}=0}^{ \substack{x''_{i,12}+x''_{j,12}+x''_{k,12} \leq \\ 2\eta''_{12} + p''_{12}}}
    \;\;
    \sum_{y''_{i,12},y''_{j,12},y''_{k,12}=0}^{ \substack{y''_{i,12}+y''_{j,12}+y''_{k,12} \leq  \\ 2\epsilon''_{12} + |m|''_{12} - p''_{12}}}
    \;\;
    \sum_{z''_{i,12},z''_{k,12}=0}^{ \substack{z''_{i,12}+z''_{k,12} \leq \\ \ell''_{12}-|m|''_{12}-2\eta''_{12}-2\epsilon''_{12}}}
    \;\;
    \sum_{x_{i,34},x_{j,34},x_{k,34}=0}^{ \substack{x_{i,34}+x_{j,34}+x_{k,34} \leq \\ 2\eta_{34} + p_{34}}}
    \;\;
    \sum_{y_{i,34},y_{j,34},y_{k,34}=0}^{ \substack{y_{i,34}+y_{j,34}+y_{k,34} \leq  \\ 2\epsilon_{34} + |m|_{34} - p_{34}}}
    \;\;
    \sum_{z_{i,34},z_{k,34}=0}^{\substack{z_{i,34}+z_{k,34} \leq \\ \ell_{34}-|m|_{34}-2\eta_{34}-2\epsilon_{34}}}
\\&
    \Bigg\{
    \frac{2^{(6 + \ell_s + f_s)(\mathrm{s_Q}-2)}}{2^{\ell_s+7}} 
    \bigg[\prod_{i \;\in \{3,4\}}
    \binom{2 \ell_{i}-2 w_{i}}{\ell_{i}-w_{i},\ell_{i}-2w_{i}-|m_{i}|,p_{i},|m_{i}|-p_{i},\eta_{i},\epsilon_{i},w_{i}-\eta_{i}-\epsilon_{i}} 
    \,\mathrm{i}^{(2\vartheta(m_{i}+\frac{1}{2})-1)(|m_{i}|-p_{i})}
    |m_{i}|!\bigg]
\\&
    (5 + \ell_s + f_s)!\;
    \mI_{\beta}(w'_1,\eta'_1,\epsilon'_1,p'_1,\ell_a,m_a,\ell_1,m_1)
    \mI_{\beta}(w'_2,\eta'_2,\epsilon'_2,p'_2,\ell_b,m_b,\ell_2,m_2)
\;
    \mI_{\delta}(f'_1,n_a,\ell_a,n_1,\ell_1)
    \mI_{\delta}(f'_2,n_b,\ell_b,n_2,\ell_2)
\\&
    (-1)^{w_s + f_s - x_{i,s} - y_{i,s} - y_{j,s}}
\;
    \mI_{\gamma}(x''_{i,12},x''_{j,12},x''_{k,12},
    2\eta'_1 + p'_1,
    2\eta'_2 + p'_2)
\;
    \mI_{\gamma}(x_{i,34},x_{j,34},x_{k,34},
    2\eta_{3} + p_{3},
    2\eta_{4} + p_{4})
\\&
    \mI_{\gamma}(y''_{i,12},y''_{j,12},y''_{k,12},
    2\epsilon'_1 + |m|'_1 - p'_1,
    2\epsilon'_2 + |m|'_2 - p'_2)
\;
    \mI_{\gamma}(y_{i,34},y_{j,34},y_{k,34},
    2\epsilon_{3} + |m_{3}| - p_{3},
    2\epsilon_{4} + |m_{4}| - p_{4})
\\&
    \mI_{\gamma}(z''_{i,12},0,z''_{k,12},
    \ell'_1-|m|'_1-2\eta'_1-2\epsilon'_1,
    \ell'_2-|m|'_2-2\eta'_2-2\epsilon'_2)
\;
    \mI_{\gamma}(t''_{12},0,0,f'_1,f'_2)
\;
    \mI_{\gamma}(t_{34},0,0,f_3,f_4)
\\&
    \mI_{\gamma}(z_{i,34},0,z_{k,34},
    \ell_{3}-|m_{3}|-2\eta_{3}-2\epsilon_{3},
    \ell_{4}-|m_{4}|-2\eta_{4}-2\epsilon_{4})
    \mS_C(
    t''_{12}, r''_{k,12},r''_{i,12},r''_{j,12},
    t_{34}, r_{k,34},r_{i,34},r_{j,34},
    p_s,|m|_s,\ell_s
    )
\\&
    \Big[
    \prod_{i \;\in \{3,4\}}
    \binom{n_i+2\ell_i+2}{n_i-f_i} \frac{1}{f_i!}
    \Big]
    \frac{
    (t''_{12} + r''_{k,12} - 1)!!
    (r''_{i,12} - 1)!!
    (r''_{j,12} - 1)!!
    (t_{34} + r_{k,34} - 1)!!
    (r_{i,34} - 1)!!
    (r_{j,34} - 1)!!
    }{
    (t''_{12} + r''_{i,12} + r''_{j,12} + r''_{k,12} + 1)!!
    (t_{34} + r_{i,34} + r_{j,34} + r_{k,34} + 1)!!
    }
\\&
    \frac{
    (x_{j,s} - y_{j,s} - p_s + 2\epsilon_s + |m|_s -1)!!
    (y_{j,s} - x_{j,s} + p_s + 2\eta_s  -1)!!
    }{
    (2\epsilon_s + 2\eta_s + |m|_s )!!
    } 
    \frac{
    (r_{i,s} + r_{j,s} + 2)!!
    (1 - r_{i,s} - r_{j,s} - r_{k,s} + \ell_s + t_s)!!
    }{
    (5 - r_{k,s} + \ell_s + t_s)!!
    } 
\\&
    \frac{
    ( 2\eta_s + 2\epsilon_s + |m|_s - x_{i,s} - x_{j,s} - y_{i,s} - y_{j,s} + z_{i,s}   )!!
    (x_{i,s} + y_{i,s} - z_{i,s} - 2\eta_s - 2\epsilon_s - |m|_s   + \ell_s -1)!!
    }{
    (1 + \ell_s - x_{j,s}  - y_{j,s})!!
    } 
    \Bigg\}\,.
\end{split}
\end{align}

\end{widetext}
In Eq.~\eqref{eq:C_result}, we have used the following auxiliary quantities that appear in the summation form of the collision integral $C$,
\allowdisplaybreaks
\begin{align}
    |m|'_1 =\;& |m_a|+|m_1|\,,
\\
    |m|'_2 =\;& |m_b|+|m_2|\,,
\\
    \ell'_1 =\;& \ell_a + \ell_1\,,
\\
    \ell'_2 =\;& \ell_b + \ell_2\,,
\\
    \xi''_{12} =\;& \xi'_1 + \xi'_2, \quad \xi\in\{|m|, \ell, f, w, p, \epsilon, \eta\}\,,
\\
    \xi_{34} =\;& \xi_3 + \xi_4, \quad \xi\in\{|m|, \ell, f, w, p, \epsilon, \eta\}\,,
\\
    \xi_{i,s} =\;& \xi''_{i,12} + \xi_{i,34}\,, \qquad \xi \in\{x,y,z\}\,,
\\
   \xi_{j,s} =\;& \xi''_{j,12} + \xi_{j,34}\,, \qquad \xi \in\{x,y\}\,,
\\
    r_{\iota,12}'' =\;& \sum_{\xi\in\{x,y,z\}} \xi_{\iota,12}''\,, \quad
    \iota\in\{i,j,k\},\; z_{j,12}'' = 0,
\\
    r_{\iota,34} =\;& \sum_{\xi\in\{x,y,z\}} \xi_{\iota,34}\,, \quad
    \iota\in\{i,j,k\},\; z_{j,34} = 0,
\\
    r_{\iota,s} =\;& r''_{\iota,12} + r_{\iota,34}\,,\qquad\iota\in\{i,j,k\}\,,
\\
    \xi_s =\;& \xi''_{12} + \xi_{34}\,, \quad \xi\in\{t, |m|, \ell, f, w, p, \epsilon, \eta\}\,.
\end{align}
The multinomial coefficient
$\binom{n}{m_1, m_2, \ldots, m_k}$
is defined in Eq.~\eqref{eq:multinomial}.
The double factorial
$n!!$
is defined in Eq.~\eqref{eq:doublefactorial}.
The selection rule factor $\mathcal{S}_C$ is a function of several intermediate variables and is given by
\begin{align}
\begin{split}
&
    \mS_C(
    t''_{12}, r''_{k,12},r''_{i,12},r''_{j,12},
    t_{34}, r_{k,34},r_{i,34},r_{j,34},
    p_s,|m|_s,\ell_s
    )
\\&=\;
    (1 + (-1)^{t''_{12} + r''_{k,12}}) 
    (1 + (-1)^{r''_{i,12}})
    (1 + (-1)^{r''_{j,12}}) 
\\&
    \times
    (1 + (-1)^{t_{34} + r_{k,34}})
    (1 + (-1)^{r_{i,34}})
    (1 + (-1)^{r_{j,34}}) 
\\&
    \times
    (1 + (-1)^{p_s}) 
    (1 + (-1)^{|m|_s}) 
    (1 + (-1)^{\ell_s} )
    \,.
\end{split}
\end{align}
We define three weight functions, $\mI_{\beta}$, $\mI_{\gamma}$, and $\mI_{\delta}$, which are defined as
\begin{align}
\begin{split}
&
    \mI_{\delta}(f'_1, n_a, \ell_a, n_1, \ell_1)
\\=\;&
    \sum_{f_a=0}^{f'_1}
    \binom{n_a+2\ell_a+2}{n_a-f_a} \frac{1}{f_a!}
    \binom{n_1+2\ell_1+2}{n_1-f_1} \frac{1}{f_1!}
    \,,
\end{split}
\end{align}
\begin{align}
\begin{split}
&
    \mI_{\beta}(w'_2,\eta'_2,\epsilon'_2,p'_2,\ell_b,m_b,\ell_2,m_2)
\\=\;&
    \binom{w'_2}{\eta'_2,\epsilon'_2,w'_2-\eta'_2-\epsilon'_2} 
    \sum_{w_b=0}^{w'_2}
    \sum_{p_b=0}^{p'_2}
    \Big[
\\\times&
    \binom{2 \ell_b-2 w_b}{w_b,\ell_b-w_b,\ell_b-2w_b-|m_b|,p_b,|m_b|-p_b} |m_b|!
\\\times&
    \binom{2 \ell_2-2 w_2}{w_2,\ell_2-w_2,\ell_2-2w_2-|m_2|,p_2,|m_2|-p_2} |m_2|!
    \Big]
\\\times&
    \,(-\mathrm{i})^{(2\vartheta(m_{b}+\frac{1}{2})-1)(|m_{b}|-p_{b})}
    \,\mathrm{i}^{(2\vartheta(m_{2}+\frac{1}{2})-1)(|m_{2}|-p_{2})}
    \;,
\end{split}
\end{align}

\begin{align}
\begin{split}
&
    \mI_{\gamma}(x,y,z,s_1,s_2)
\\=\;&
    \sum_{x_1=0}^{x}
    \sum_{y_1=0}^{y}
    \sum_{z_1=0}^{z}
    (-1)^{x_1+y_1+z_1}
\\\times&
    \Big[\hatprod_{i=1,2}
    \binom{s_i}{x_i,y_i,z_i,s_i-x_i-y_i-z_i}
    \Big]
    \;,
\end{split}
\end{align}
where $w_2 = w'_2-w_b$, $p_2 = p'_2-p_b$, $x_2=x-x_1$,$y_2=y-y_1,z_2=z-z_1$, and $f_1 = f'_1 - f_a$.

\begin{widetext}
To compute $\tilde{A}_{ijksq}$, it suffices to evaluate the case $m_i = m_j = m_k = m_s = m_q = 0$; all other components can be systematically reconstructed using the properties discussed in Sec.~\ref{section:ApplyRatio}.
The final result for $\tilde{A}_{ijksq}$ is
\begin{align}
\begin{split}
&
    A_{n_a \ell_a 0, n_1 \ell_1 0, n_2 \ell_2 0, n_3 \ell_3 0, n_4 \ell_4 0 }
\\=\;&
    \frac{n_a!}{(2l_a+n_a+2)!}
    \frac{\sigma^{(0)}\Lambda}{2^{\ell_s+7} \pi^\frac{9}{2}} 
    \bigg[\prod_{i\,\in\{a,1,2,3,4\}}
    \sqrt{2 \ell_i+1}\bigg]
    \sum_{w'_{1}=0}^{
    \lfloor\frac{\ell_a}{2}\rfloor
    +
    \lfloor\frac{\ell_{1}}{2}\rfloor}
    \sum_{\eta'_{1}=0}^{w'_{1}} 
    \sum_{\epsilon'_{1}=0}^{w'_{1}-\eta'_{1}}
    \sum_{f'_1=0}^{n_a+n_1}
    \bigg[
    \hatprod_{i \;\in \{2,3,4\}}
    \sum_{w_{i}=0}^{\lfloor\frac{\ell_{i}}{2}\rfloor}
    \sum_{\eta_{i}=0}^{w_{i}} 
    \sum_{\epsilon_{i}=0}^{w_{i}-\eta_{i}}
    \sum_{f_i=0}^{n_i}
    \bigg]
    \sum_{t'_{12}=0}^{f'_1+f_2}
\\&
    \sum_{t_{34}=0}^{f_3+f_4}
    \sum_{x'_{i,12},x'_{j,12},x'_{k,12}=0}^{\substack{x'_{i,12}+x'_{j,12}+x'_{k,12} \leq \\ 2\eta{'}_{12}}}
    \quad
    \sum_{y'_{i,12},y'_{j,12},y'_{k,12}=0}^{\substack{y'_{i,12}+y'_{j,12}+y'_{k,12} \leq  \\ 2\epsilon{'}_{12}}}
    \quad
    \sum_{z'_{i,12},z'_{k,12}=0}^{\substack{z'_{i,12}+z'_{k,12} \leq \\ \ell{'}_{12}-2\eta{'}_{12}-2\epsilon{'}_{12}}}
    \sum_{x_{i,34},x_{j,34},x_{k,34}=0}^{\substack{x_{i,34}+x_{j,34}+x_{k,34} \leq \\ 2\eta_{34}}}
    \quad
    \sum_{y_{i,34},y_{j,34},y_{k,34}=0}^{\substack{y_{i,34}+y_{j,34}+y_{k,34} \leq  \\ 2\epsilon_{34}}}
    \quad
    \sum_{z_{i,34},z_{k,34}=0}^{\substack{z_{i,34}+z_{k,34} \leq \\ \ell_{34}-2\eta_{34}-2\epsilon_{34}}}
\\&
    \Bigg\{
    (5 + \ell_s + f_s)!
    \frac{(-1)^{w_s + f_s - x_{i,s} - y_{i,s} - y_{j,s}}}
    {2^{(6 + \ell_s + f_s)(2-\mathrm{s_Q})}}
    \bigg[\prod_{i\;\in\{2,3,4\}}
    \binom{2 \ell_i-2 w_i}{\ell_i-w_i,\ell_i-2w_i,\eta_i,\epsilon_i,w_i-\eta_i-\epsilon_i}
    \binom{n_i+2\ell_i+2}{n_i-f_i} \frac{1}{f_i!}
    \bigg]
\\&
    \mI_{\gamma}(x'_{i,12},x'_{j,12},x'_{k,12},
    2\eta'_1,
    2\eta_2)
\;
    \mI_{\gamma}(y'_{i,12},y'_{j,12},y'_{k,12},
    2\epsilon'_1,
    2\epsilon_2)
\;
    \mI_{\gamma}(z'_{i,12},0,z'_{k,12},
    \ell'_1-2\eta'_1-2\epsilon'_1,
    \ell_2-2\eta_2-2\epsilon_2)
\\&
    \mI_{\gamma}(x_{i,34},x_{j,34},x_{k,34},
    2\eta_{3},
    2\eta_{4})
\;
    \mI_{\gamma}(y_{i,34},y_{j,34},y_{k,34},
    2\epsilon_{3},
    2\epsilon_{4})
\;
    \mI_{\gamma}(z_{i,34},0,z_{k,34},
    \ell_{3}-2\eta_{3}-2\epsilon_{3},
    \ell_{4}-2\eta_{4}-2\epsilon_{4})
\\&
    \mI_{\delta}(f'_1,n_a,\ell_a,n_1,\ell_1)
\;
    \mI_{\gamma}(t{'}_{12},0,0,f'_1,f_2)\;
    \mI_{\gamma}(t_{34},0,0,f_3,f_4)
\;
    \mS_A(
    t'_{12}, r'_{k,12},r'_{i,12},r'_{j,12},
    t_{34}, r_{k,34},r_{i,34},r_{j,34},
    \ell_s
    )
\\&
    \mI_{\beta}(w'_1,\eta'_1,\epsilon'_1,0,\ell_a,0,\ell_1,0)
    \frac{
    (t'_{12} + r'_{k,12} - 1)!!
    (r'_{i,12} - 1)!!
    (r'_{j,12} - 1)!!
    }{
    (t'_{12} + r'_{i,12} + r'_{j,12} + r'_{k,12} + 1)!!
    }
    \frac{
    (t_{34} + r_{k,34} - 1)!!
    (r_{i,34} - 1)!!
    (r_{j,34} - 1)!!
    }{
    (t_{34} + r_{i,34} + r_{j,34} + r_{k,34} + 1)!!
    }
\\&
    \frac{
    (r_{i,s} + r_{j,s} + 2)!!
    (1 - r_{i,s} - r_{j,s} - r_{k,s} + \ell_s + t_s)!!
    }{
    (5 - r_{k,s} + \ell_s + t_s)!!
    } 
    \frac{
    (x_{j,s} - y_{j,s} + 2\epsilon_s -1)!!
    (y_{j,s} - x_{j,s} + 2\eta_s-1)!!
    }{
    (2\epsilon_s + 2\eta_s)!!
    } 
\\&
    \frac{
    (z_{i,s}  - x_{i,s} - x_{j,s} - y_{i,s} - y_{j,s} + 2\eta_s  + 2\epsilon_s)!!
    (x_{i,s} + y_{i,s} - z_{i,s} - 2\eta_s -2\epsilon_s  + \ell_s -1)!!
    }{
    (1 + \ell_s - x_{j,s}  - y_{j,s})!!
    }
    \Bigg\}\,,
\end{split}\label{eq:A_result}
\end{align}
\end{widetext}
where
\allowdisplaybreaks
\begin{align}
    \ell'_1 =\;& \ell_a + \ell_1\,,
\\
    \xi'_{12} =\;& \xi'_1 + \xi_2, \quad \xi\in\{ \ell, f, w, \epsilon, \eta\}\,,
\\
    \xi_{34} =\;& \xi_3 + \xi_4, \quad \xi\in\{ \ell, f, w, \epsilon, \eta\}\,,
\\
    \xi_s =\;& \xi{'}_{12} + \xi_{34}\,, \quad \xi\in\{t, \ell, f, w, \epsilon, \eta\}\,,
\\
    \xi_{i,s} =\;& \xi'_{i,12} + \xi_{i,34}\,, \qquad \xi \in\{x,y,z\}\,,
\\
   \xi_{j,s} =\;& \xi'_{j,12} + \xi_{j,34}\,, \qquad \xi \in\{x,y\}\,,
\\
    r_{\iota,12}' =\;& \sum_{\xi\in\{x,y,z\}} \xi_{\iota,12}'\,, \quad
    \iota\in\{i,j,k\},\; z_{j,12}' = 0,
\\
    r_{\iota,34} =\;& \sum_{\xi\in\{x,y,z\}} \xi_{\iota,34}\,, \quad
    \iota\in\{i,j,k\},\; z_{j,34} = 0,
\\
    r_{\iota,s} =\;& r'_{\iota,12} + r_{\iota,34}\,,\qquad\iota\in\{i,j,k\}\,.
\end{align}
The selection rule factor $\mathcal{S}_A$ takes the form
\begin{align}
\begin{split}
&
    \mS_A(
    t'_{12}, r'_{k,12},r'_{i,12},r'_{j,12},
    t_{34}, r_{k,34},r_{i,34},r_{j,34},
    \ell_s
    )
\\=\;&
    (1 + (-1)^{t'_{12} + r'_{k,12}}) 
    (1 + (-1)^{r'_{i,12}})
    (1 + (-1)^{r'_{j,12}}) 
\\\times&
    (1 + (-1)^{t_{34} + r_{k,34}})
    (1 + (-1)^{r_{i,34}})
    (1 + (-1)^{r_{j,34}}) 
\\\times&
    (1 + (-1)^{\ell_s} ).
\end{split}
\end{align}

\section{Numerical Validation}\label{sec:Result}

From an analytical perspective, the spectral BBGKY hierarchy is exactly equivalent to the conventional BBGKY hierarchy. This equivalence relies on expanding the reduced distribution functions in a complete set of basis functions in momentum space. In practical numerical implementations, however, the expansion must be truncated to a finite number of modes. It is therefore essential to demonstrate that the truncated spectral BBGKY hierarchy still provides a well-posed and accurate approximation. 
We therefore validate the truncation in the present section.

Our assessment of the accuracy of the spectral BBGKY hierarchy is based on the analytic evaluation of the collision kernel presented in Sec.~\ref{subsec:Summation Form of the Collision Kernel}. Specifically, we consider the case of massless particles with a transition probability specified by Eq.~\eqref{eq:example_collision_kernel}. 
The collision integral associated with the transition probability in Eq.~\eqref{eq:example_collision_kernel} is shown in Eqs.~\eqref{eq:C_result} and \eqref{eq:A_result}.

In this section, we demonstrate that, with the choice of basis functions in Eq.~\eqref{eq:basis}, only the first few low-order components are sufficient to accurately approximate the nonlinear evolution of the distribution function.
We present four types of evidence to support the validity of the spectral method:
(i) strict preservation of conservation laws,
(ii) agreement with a known analytical solution,
(iii) numerical convergence over the spectral expansion, and
(iv) negligible leakage of expansion coefficients.
These results provide a solid foundation for truncating the hierarchy in practice, making the spectral method both efficient and reliable for numerical studies.
We further provide recommendations for selecting the truncation parameters $(n_{\mathrm{max}}, \ell_{\mathrm{max}})$.

\vspace{3mm}
The most fundamental requirement for any kinetic theory approach is the preservation of conserved quantities. 
Under our choice of basis in Eq.~\eqref{eq:basis}, the five essential conserved quantities---particle number, energy, and the three components of momentum—are exactly preserved. 
In Eqs.~(\ref{eq:conservationAndCoefficient_1}-\ref{eq:conservationAndCoefficient_5}), we show how these conserved quantities correspond to specific expansion coefficients. 
These specific expansion coefficients, $f_{i_{\mathrm{conserved}}}$, do not evolve under collisions because of the antisymmetric nature of the corresponding collision matrix $A_{i_{\mathrm{conserved}}jk}$. 
As a result, the numerical evolution strictly respects conservation laws.

\begin{figure}
    \centering
    \includegraphics[width=1\linewidth]{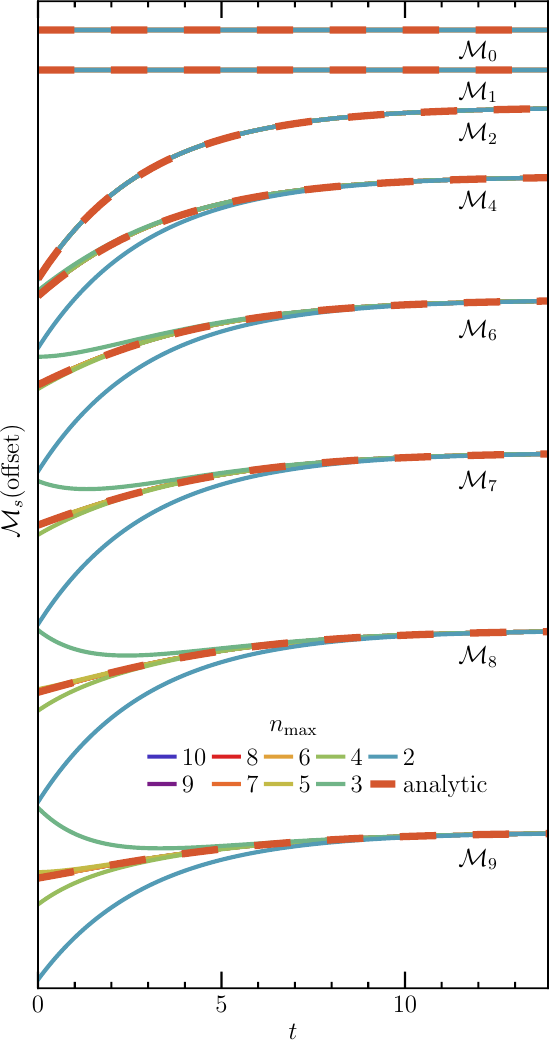}
    \caption{\textbf{Comparison between spectral and analytical results across various truncation levels.} Numerical results for truncation orders $n_\mathrm{max} = 2, 3, \dots, 10$ (thin rainbow-colored solid lines) are compared against the analytical solution (thick orange dashed line), demonstrating the convergence behavior of the spectral method.
    For visual clarity, each moment ($\mathcal{M}_s$) is vertically offset and rescaled.
    In many panels, the curve for higher $n_{\mathrm{max}}$ is visually indistinguishable, as it is completely overlapped by that of a lower truncation.
    The top two curves, $\mathcal{M}_0$ and $\mathcal{M}_1$, represent particle number and energy density. Their time-independence reflects the exact conservation of these quantities.}
    \label{fig:compare2016PRL}
\end{figure}

\vspace{3mm}
Next, we validate the spectral method by comparing our numerical calculations with a known analytical solution of the nonlinear Boltzmann equation found in Ref.~\cite{Bazow:2015dha}.
The same assumptions are adopted as in this section, namely: (i) the particles are massless, and (ii) the transition probability takes the form of Eq.\eqref{eq:example_collision_kernel}.
The analytical solution starts from a specific initial condition that is homogeneous in coordinate space and isotropic in momentum,
\begin{align}
f(t=0,\bp)=\frac{256}{243} \frac{|\bp|}{T_0}  e^{- \frac{4 |\bp|}{3 T_0}}\,.
\end{align}
We simulate the system under various truncation levels, keeping $\ell_{\mathrm{max}} = 0$ fixed due to isotropy and varying $n_{\mathrm{max}} = 2, 3, \ldots, 10$. To quantify the accuracy, we track the evolution of the moments of energy ($E = p^0$),
\begin{align}
\mathcal{M}_s(t) = \int d^3 \bp \, E^s f(t, \boldsymbol{p})\,,
\end{align}
and compare them to the analytical results, as shown in Fig.~\ref{fig:compare2016PRL}.
To highlight qualitative features across different moments, each moment is displayed with an individual shift and rescaling.
The curves corresponding to $\mathcal{M}_0$ and $\mathcal{M}_1$ represent the evolution of the particle number density and energy density, respectively. As expected, both quantities remain constant over time, reflecting the exact conservation laws encoded in the simulation.
For other moments, excellent agreement is obtained for low-order moments across all truncation levels. As expected, high-order moments deviate under small truncations, but this deviation decreases rapidly with increasing $n_{\mathrm{max}}$, indicating fast convergence of the basis expansion. Specifically, truncation levels $n_{\mathrm{max}} = 2$, 3, 4, 5, and 6 are sufficient to accurately reproduce the moments $\mathcal{M}_2$, $\mathcal{M}_4$, $\mathcal{M}_6$, $\mathcal{M}_8$, and $\mathcal{M}_9$, respectively.
The excellent agreement with the analytical solution observed in Fig.~\ref{fig:compare2016PRL} serves to demonstrate the robustness of the spectral BBGKY method.

\begin{figure*}
    \centering
    \includegraphics[width=1\textwidth]{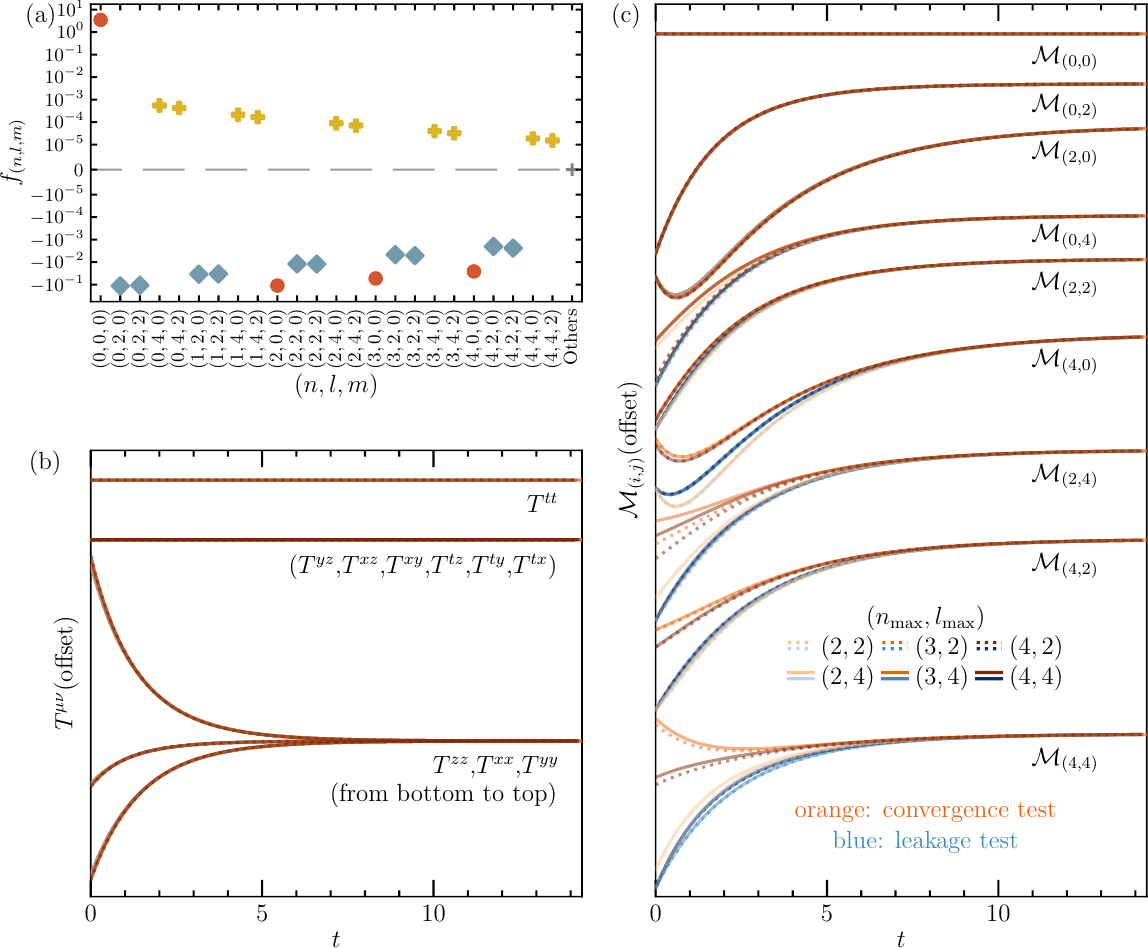}
    \caption{
    \textbf{Numerical convergence and leakage tests of the spectral BBGKY method, under a representative anisotropic initial condition as defined in Eq.~\eqref{eq:anisotropic_initial_state}.}
    (a) Decomposition of the initial distribution function into spectral components. 
    (b) Time evolution of the components of the energy-momentum tensor $T^{\mu\nu}$.
    (c) Time evolution of selected moments $\mathcal{M}_{(i,j)}$ under various truncation schemes.
    In panels (b) and (c), orange curves (with varying shades and line styles) depict the nonlinear evolution of the initial state under different truncation levels, $(n_{\mathrm{max}}, \ell_{\mathrm{max}}) = (2,2), (2,4), \ldots,(4,4)$.
    Blue curves correspond to the evolution of the initial state truncated at $n \le 2$, $\ell \le 2$, but using larger bases during time evolution, $(n_{\mathrm{max}}, \ell_{\mathrm{max}}) = (2,4)$ (light dashed), $(3,2)$ (medium dashed), $(3,4)$ (medium solid), $(4,2)$ (dark dashed), and $(4,4)$ (dark solid).
    In the vertical direction, offsets and distinct scales are applied to $T^{\mu\nu}$ and $\mathcal{M}_{(i,j)}$ curves to enhance visual clarity.
    The conservation of particle number density, energy density, and momentum density is manifested by the time-invariant behavior of $\mathcal{M}_{(0,0)}$, $T^{tt}$, and $T^{tx}$, $T^{ty}$, $T^{tz}$, respectively.
    }
    \label{fig:converge_physics_initial_state}
\end{figure*}

We also check the convergence of the spectral method by considering a representative anisotropic initial condition,
\begin{align}\label{eq:anisotropic_initial_state}
\begin{split}
    f(t=0,\bp)
=\;&
    \frac{256}{243} \frac{|\bp|}{T_0} 
    \exp\left(- \frac{4 |\bp|\sqrt{[1+(\xi-1)\cos^2 \theta]}}{3 T_0}\right)
\\&
    \times
    (1+2 v_2 \cos (2\phi))
    \,,
\end{split}
\end{align}
where $\theta$ and $\phi$ denote the polar and azimuthal angles of the momentum vector $\bp$ in spherical coordinates, respectively. Here, $\xi$ and $v_2$ characterize the longitudinal and transverse degrees of momentum anisotropy, respectively.
We study the nonlinear evolution under several truncation levels, 
\begin{align}\label{eq:truncate_physics}
    (n_{\mathrm{max}}, \ell_{\mathrm{max}}) = (2,2), (2,4), (3,2), (3,4), (4,2), (4,4)
    \,,
\end{align}
keeping all magnetic components $-\ell \leq m \leq \ell$.
The analysis is limited to even values of $\ell_{\mathrm{max}}$ because of the parity symmetry of the initial state.
To quantify the accuracy, we track the evolution of the moments
\begin{align}
\mathcal{M}_{(i,j)}(t) = \int d^3 \bp \, E^i p_z^j f(t, \boldsymbol{p})\,,
\end{align}
and the evolution of the energy-momentum tensor
\begin{align}
    T^{\mu\nu}
=\;&
    \int \frac{d^3 \bp}{p^0} p^\mu p^\nu f(t,\bp)
    \,.
\end{align}

Results are shown in Fig.~\ref{fig:converge_physics_initial_state}, where a highly anisotropic initial state characterized by $\xi = 10$ and $v_2 = -\frac{1}{2}$ is employed. The corresponding spectral coefficients are displayed in panel (a). As expected for a physical distribution, they decay rapidly with increasing $n$ and $\ell$, which ensures the applicability of the spectral method.
We present numerical results for both the energy-momentum tensor $T^{\mu\nu}$ in panel (b) and for the moments in panel (c), represented by orange curves with varying shades and line styles.
To aid visual interpretation, curves in both panels (b) and (c) are shown with offsets and individual rescaling along the vertical axis. Only qualitative behavior across different observables is intended to be compared.
\begin{figure*}[!hbtp]
    \centering
    \includegraphics[width=1.0\textwidth]{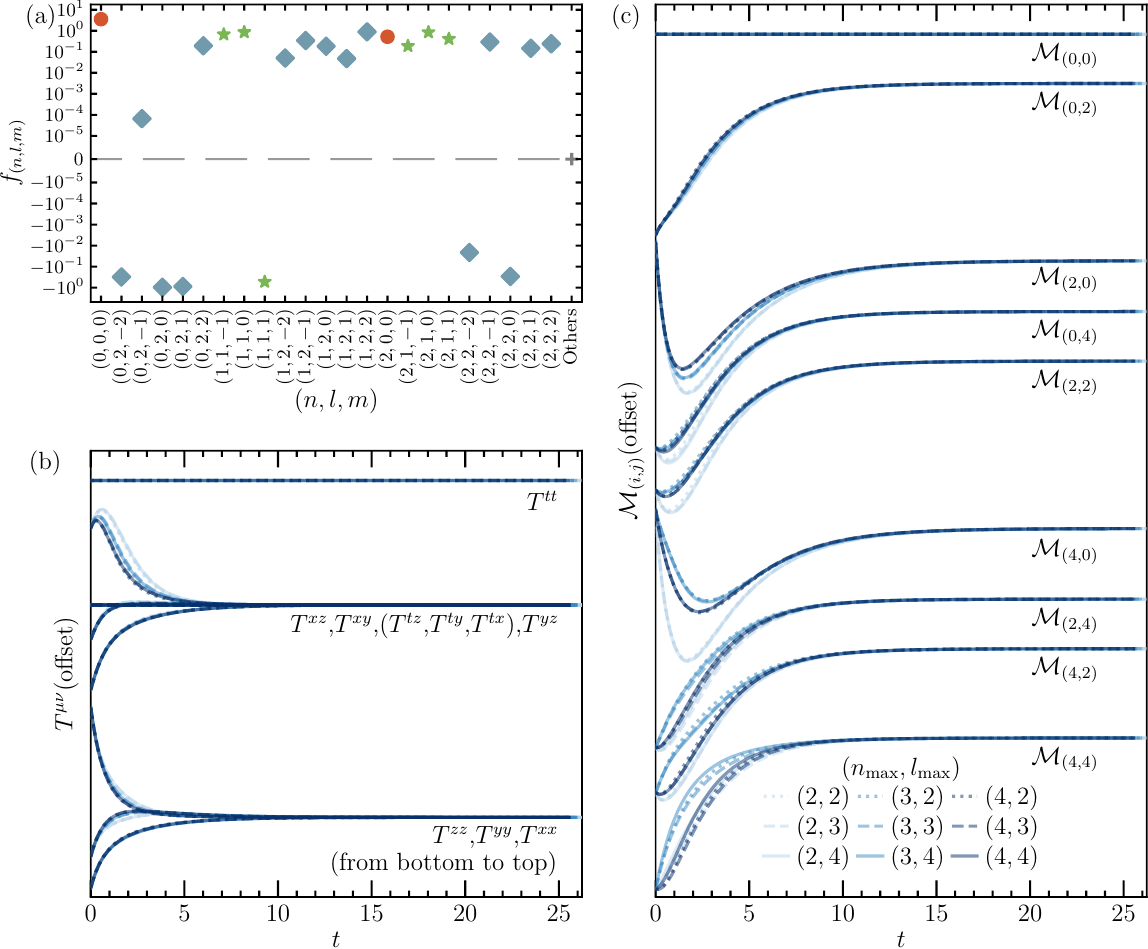}
    \caption{
    \textbf{Leakage tests under a random initial condition generated from Eq.~\eqref{eq:truncate_random}.} 
    Conventions and settings are the same as in Fig.~\ref{fig:converge_physics_initial_state}.  
    }
    \label{fig:converge_random_initial_state}
\end{figure*}

In panel (b), the components of the energy-momentum tensor $T^{\mu\nu}$ are grouped into three distinct categories based on their physical behavior. 
Different components in the same group share the same offset and rescaling constants. 
The first group consists of $T^{tt}$, which represents the energy density and remains a nonzero constant throughout the evolution, reflecting the conservation of energy density. 
The second group comprises the off-diagonal components $T^{tz}$, $T^{ty}$, $T^{tx}$, $T^{yz}$, $T^{xz}$, and $T^{xy}$, all of which remain zero over time. 
In particular, $T^{tz}$, $T^{ty}$, and $T^{tx}$ remain zero not only due to a manifestation of momentum density conservation, but also because the system is initialized in the local rest frame, where the net momentum is zero. 
The remaining components, $T^{yz}$, $T^{xz}$, and $T^{xy}$, remain zero due to the initial isotropy of the distribution. 
The third group includes the diagonal spatial components $T^{xx}$, $T^{yy}$, and $T^{zz}$, whose time evolution captures the isotropization process of the system. 
Notably, at late times, these components converge to an identical constant values, indicating the restoration of isotropy in the spatial stress tensor.
We find that curves corresponding to the same moment coincide with each other in the energy-momentum tensor, indicating that $T^{\mu\nu}$ can already be described accurately with $(n_{\mathrm{max}}, \ell_{\mathrm{max}}) = (2,2)$. 
This demonstrates that only a small basis set is required to accurately capture the hydrodynamic evolution of the system.

In addition, trends similar to those in Fig.~\ref{fig:compare2016PRL} are observed in panel (c).
The constant behavior of the moment $\mathcal{M}_{(0,0)}$ over time reflects the exact conservation of particle number density.
The solutions exhibit convergence up to $n_\mathrm{max} = 4$, as indicated by the smaller discrepancy between the $n_\mathrm{max} = 3$ and $n_\mathrm{max} = 4$ curves compared to that between $n_\mathrm{max} = 2$ and $n_\mathrm{max} = 3$.
We also observe minor differences between the dashed and solid lines of the same color, which correspond to $\ell_{\mathrm{max}} = 2$ and $4$, respectively, for the same $n_{\mathrm{max}}$. This indicates convergence with respect to the angular expansion parameter $\ell_{\mathrm{max}}$.

\vspace{3mm}
As a reference case, we perform a simulation in which the initial spectral coefficients $f_{(n,\ell,m)}$ are set to zero for $n > 2$ or $\ell > 2$, while the remaining coefficients are initialized according to Eq.~\eqref{eq:anisotropic_initial_state}. We then evolve the distribution using larger basis sets, and the results are shown in blue.

We observe that all blue curves are very close to each other, and they are all in good agreement with the orange curves corresponding to $(n_{\mathrm{max}}, \ell_{\mathrm{max}}) = (2,2)$. For many moments, the blue curves are almost entirely overlapped by the orange curves. This comparison shows that the higher-order $f_{(n,\ell,m)}$ coefficients remain small if they are initialized as zero.
From this, we conclude that the leakage of coefficients due to nonlinear coupling from low-order to high-order modes is negligible when the expansion coefficients remain small. This provides further support for the validity of truncating the hierarchy in practical simulations.

Finally, we repeat the leakage analysis for a randomly sampled initial state,
\begin{align}\label{eq:truncate_random}
\begin{cases}
f_{(n,\ell,m)} =  2\sqrt{\pi} &n=0,\; \ell=0, \\
f_{(n,\ell,m)} \sim \mathcal{U} &
    (n,\ell) = (0,2), (1,1), \text{or}~(1,2),\\
f_{(n,\ell,m)} \sim \mathcal{U} &
    n=2,\;\ell\leq 2,\\
f_{(n,\ell,m)} = 0 & \mathrm{otherwise,}
\end{cases}
\end{align}
where $\mathcal{U}$ denotes a uniform distribution defined on the interval $[-1,1]$.
The sampled initial coefficients are shown in Fig.~\ref{fig:converge_random_initial_state}~(a).
The results of the leakage test, viewed from the perspectives of the energy-momentum tensor $T^{\mu\nu}$ and the moments $\mathcal{M}_{(i,j)}$, are presented in panels (b) and (c), respectively.
For clearer visualization, in both panels (b) and (c), curves are plotted with shifted vertical positions and varying scales.

Similar to Fig.~\ref{fig:converge_physics_initial_state}~(b), the components of the energy-momentum tensor $T^{\mu\nu}$ are organized into three categories. 
The invariant behavior of $T^{tt}$ reflects the conservation of energy density, while the constancy of $T^{tx}$, $T^{ty}$, and $T^{tz}$ indicates momentum conservation. 
For the random initial conditions considered here, the off-diagonal components $T^{yz}$, $T^{xz}$, and $T^{xy}$ are initially nonzero but decay to zero at late times, signaling the suppression of anisotropic stress. 
As in Fig.~\ref{fig:converge_physics_initial_state}~(b), the diagonal spatial components $T^{xx}$, $T^{yy}$, and $T^{zz}$ evolve toward a common asymptotic value, marking the restoration of isotropy in the spatial stress tensor.

For the energy-momentum tensor shown in Fig.~\ref{fig:converge_random_initial_state}~(b),
a slightly larger discrepancy is observed between the leakage test curves (blue lines) compared to those in Fig.~\ref{fig:converge_physics_initial_state}~(b).
Nonetheless, the differences between truncations with the same $\ell_{\mathrm{max}}$ (i.e., curves with the same color but different line styles) remain nearly negligible.
This suggests that even for an initial distribution characterized by a Heaviside-like structure in its expansion coefficients, a truncation at $\ell_{\mathrm{max}} = 2$ is already sufficient to capture the hydro-relevant angular dynamics.
The main variation arises from truncations with the same $\ell_{\mathrm{max}}$ but different $n_{\mathrm{max}}$ values (i.e., curves with the same line style but different colors).
Notably, the difference between the results for $n_{\mathrm{max}} = 4$ and $n_{\mathrm{max}} = 3$ is smaller than that between $n_{\mathrm{max}} = 3$ and $n_{\mathrm{max}} = 2$.
This indicates that, even when the initial expansion coefficients exhibit a sharp step-like behavior, accurate evolution can still be achieved by selecting a truncation slightly above the step location.

In panel (c), the moment $\mathcal{M}_{(0,0)}$ exhibits no temporal variation, indicating the conservation of particle number density. 
The next five moments---$\mathcal{M}_{(0,2)}$, $\mathcal{M}_{(2,0)}$, $\mathcal{M}_{(0,4)}$, $\mathcal{M}_{(2,2)}$, and $\mathcal{M}_{(4,0)}$---exhibit convergence behavior consistent with that observed in panel (b). 
Notably, the evolution trajectories for a fixed $n_{\mathrm{max}}$ remain virtually unchanged as $\ell_{\mathrm{max}}$ increases from 2 to 4, indicating that $\ell_{\mathrm{max}} = 2$ is already sufficient to resolve the angular structure of these low-order modes.
Additionally, a clear convergence trend is seen with respect to $n_{\mathrm{max}}$.
The discrepancy between $n_{\mathrm{max}} = 4$ and $n_{\mathrm{max}} = 3$ is significantly smaller than that between $n_{\mathrm{max}} = 3$ and $n_{\mathrm{max}} = 2$.
This reinforces the conclusion that, even when the initial spectral coefficients exhibit sharp features---such as step-like discontinuities---the evolution of moments can still be accurately captured. 
For lower-order moments, reliable results can be achieved using moderate truncation parameters, 
Specifically, $\ell_{\mathrm{max}} = 2$ and $n_{\mathrm{max}}$ chosen just beyond the dominant mode present in the initial distribution.

In contrast, the behavior of higher-order moments---such as $\mathcal{M}_{(2,4)}$, $\mathcal{M}_{(4,2)}$, and $\mathcal{M}_{(4,4)}$---reveals a greater sensitivity to the truncation in the angular quantum number. 
In addition, for these moments, the differences between the results with $n_{\mathrm{max}} = 3$ and $n_{\mathrm{max}} = 4$ are no longer smaller than those between $n_{\mathrm{max}} = 2$ and $n_{\mathrm{max}} = 3$. 
This indicates that higher truncation parameters, $(n_{\mathrm{max}}, \ell_{\mathrm{max}})$, are needed to accurately capture the evolution of higher-order moments, when the initial coefficients exhibit a Heaviside-like structure across different basis functions.

The results shown in Fig.~\ref{fig:converge_physics_initial_state} and Fig.~\ref{fig:converge_random_initial_state} further corroborate the robustness of the spectral BBGKY method, as demonstrated by the convergence and leakage tests.

\section{Summary and Outlook} \label{sec:Discussion and Conclusion}

In this work, we proposed the spectral BBGKY hierarchy, a reformulation of the conventional BBGKY hierarchy that maintains full analytical equivalence while offering significant advantages in numerical implementation. This framework is particularly suitable for studying the full nonlinear Boltzmann equation and multi-particle nonlinear correlations in high-dimensional phase space, where traditional methods often face severe computational limitations.

By decomposing the reduced distribution functions into a complete orthonormal basis [Eq.~\eqref{eq:basis}], the angular dependence of the many-body system is encoded via spherical harmonics. This decomposition enables a transparent description of the isotropization process. 
Moreover, with an appropriate choice of the temperature scale parameter $\Lambda$, the equilibrium distribution takes on a simple analytical form [Eq.~\eqref{eq:coef_thermal_special}]. 
The radial basis functions are constructed from generalized Laguerre polynomials to ensure exact enforcement of the five conservation laws---particle number, energy, and the three components of momentum [Eqs.~(\ref{eq:conservationAndCoefficient_1}-\ref{eq:conservationAndCoefficient_5})]---even at low truncation orders.

Beyond these advantages from a physical perspective, the spectral BBGKY hierarchy eliminates the need to discretize momentum space.
Instead, the momentum dependence of the distribution function is encoded in a set of spectral coefficients.
By replacing direct discretization of momentum space with a spectral expansion over $M$ modes characterizing the single-particle momentum dependence (typically $M=27$), the spectral BBGKY hierarchy shifts the computational cost from discretizing a $6n$-dimensional phase space to evolving $M^n$ spectral coefficients over a $3n$-dimensional spatial domain. 
Given that $M$ is significantly smaller than the typical number of grid points used per momentum dimension (e.g., 100), the resulting memory and computational savings are substantial.
In addition to memory savings, spectral methods can achieve significantly higher accuracy than grid-based discretization schemes, provided that appropriate basis functions are chosen. Spectral methods exhibit exponential convergence when applied to sufficiently smooth problems, with the numerical error scaling as $O(e^{-M})$, where $M$ denotes the number of basis functions. In contrast, grid-based methods generally display algebraic convergence, with errors of the order $O(\Delta_p^n)$, where $\Delta_p$ is the momentum grid spacing and $n$ is the numerical order of accuracy for integration or differentiation---commonly $n = 2$.
Restricting to a typical single particle distribution function considered in the Boltzmann equation, i.e., $n=1$,
the numerical resource requirement is comparable to that of the Lattice Boltzmann Method (LBM). However, unlike LBM, which evolves a linearized version of the Boltzmann equation, our method enables the evolution of the full nonlinear Boltzmann equation at a similar computational cost. 
When extended to higher-order truncations (i.e., $n \ge 2$), this dimensional reduction renders the method both scalable and well-suited for studying the nonlinear evolution of multi-particle correlations.
The aforementioned simplification is applicable not only to elastic scattering [Eq.~\eqref{eq:short_interaction}] but also to inelastic processes, including both $2 \leftrightarrow 2$ and $2 \leftrightarrow 3$ collisions, with only minor modifications required in the collision term of Eq.~\eqref{eq:short_interaction}. 
Although the interaction considered in this work is short-range, long-range interactions can also be incorporated by introducing gauge fields. For instance, electromagnetic interactions may be treated by including photons as an explicit degree of freedom in the distribution function.

We have also developed an analytic method for evaluating the collision integrals by expanding the differential cross section in its own orthogonal basis [Eq.~\eqref{eq:sigma_legendre_expansion}]. This approach evaluates the full eight-dimensional integral exactly for massless particles, and reduces it to a three-dimensional integral for massive particles. It removes the need for ensemble averaging over repeated stochastic evolutions from the same initial state, as is typically required in geometrical or stochastic cascade methods. The nonlinear evolution can instead be accurately captured by a single deterministic simulation.

We have also performed convergence and leakage analyses based on the proposed spectral BBGKY hierarchy and the analytic evaluation of the collision integrals. 
As expected, key analytic properties---such as exact conservation laws and the structure of thermal equilibrium---are consistently recovered, reinforcing the method's reliability and physical consistency. 
The results also demonstrate that the nonlinear dynamics of the system can be accurately captured using a modest set of low-order modes. For a representative single-particle distribution function [Eq.~\eqref{eq:anisotropic_initial_state}], we show that a truncation at $(n_{\mathrm{max}}, \ell_{\mathrm{max}}) = (2,2)$ is sufficient to resolve the nonlinear evolution of the energy-momentum tensor $T^{\mu\nu}$---the key hydrodynamic observable. To accurately describe the evolution of higher-order moments $\mathcal{M}_{(i,j)}$, a truncation of $(n_{\mathrm{max}}, \ell_{\mathrm{max}}) = (4,2)$ is found to be adequate. These analyses also offer practical guidance for the choice of truncation parameters $(n_{\mathrm{max}}, \ell_{\mathrm{max}})$.

Overall, the spectral BBGKY hierarchy offers an analytically rigorous and computationally scalable reformulation of the conventional BBGKY hierarchy. At the lowest truncation level, to the best of our knowledge, it provides the most efficient and accurate approach for solving the full nonlinear Boltzmann equation. Moreover, it enables systematic, high-precision nonlinear simulations of multi-particle correlations far from equilibrium. 
This method can be used to explore non-equilibrium dynamics across a broad range of physical systems.

This spectral framework opens up a range of directions for future investigation in kinetic theory and non-equilibrium physics.
First, within the conventional Boltzmann framework, once a system reaches thermal equilibrium, its macroscopic properties can be fully characterized by a few thermodynamic variables---such as temperature, volume, and particle number density. All other information, including details of the initial conditions and interactions (unless tied to conserved quantities), is lost.
Whether this erasure of microscopic information remains valid when particle correlations are retained is a fundamental question that deserves further exploration.
Second, for systems in which correlations are non-negligible, the spectral BBGKY hierarchy may offer new insights. In particular, problems such as early thermalization in high-energy nuclear collisions could benefit from a correlation-based perspective.
Third, our current formulation assumes a Boltzmann equilibrium distribution, which neglects quantum statistical factors, and applies analytic collision integral evaluation only to massless particles. Extending this approach to include massive particles would be a natural next step. Furthermore, developing analogous spectral frameworks for quantum-statistical systems, such as those obeying Bose--Einstein or Fermi--Dirac statistics, is also a promising direction for future research.
\section*{Acknowledgments} 
We are grateful to Baoyi Chen, Jin Hu, Anping Huang, Jiaxin Luo, Kaijia Sun, Zhe Xu, Xingbo Zhao, Fabian Zhou, Tianzhe Zhou, and Pengfei Zhuang for helpful discussions.
This work is supported by Tsinghua University under grant Nos. 04200500123, 531205006, 533305009.
We also acknowledge the support by center of high performance computing, Tsinghua University.

\appendix
\section{Spherical Harmonics}\label{Appendix:Spherical Harmonics}

\subsection{Complex spherical harmonics $Y_{\ell}^m: S^2 \rightarrow \mathbb{C}$}

\textbf{Conventions}

The complex spherical harmonics $Y_{\ell}^m(\theta, \varphi)$ are defined as  
\begin{align}
\begin{split}
Y_{\ell}^{m}(\theta, \varphi)=
\sqrt{\frac{(2 \ell+1)}{4 \pi} \frac{(\ell-m)!}{(\ell+m)!}} P_{\ell}^m(\cos \theta) e^{\mathrm{i} m \varphi},
\\
\quad \text{where} \quad -\ell \leq m \leq \ell \,,
\end{split}
\end{align}  
where $P_{\ell}^m(x)$ are the associated Legendre polynomials, defined by
\begin{align}
P_{\ell}^m(x) = \frac{(-1)^m}{2^{\ell} \ell!} (1-x^2)^{m / 2} \frac{d^{\ell+m}}{d x^{\ell+m}} (x^2-1)^{\ell} \,.
\end{align}  

\textbf{Conjugate Symmetry}

The complex spherical harmonics satisfy the following conjugate symmetry relation
\begin{align}
Y_{\ell}^{-m}(\theta, \phi)=(-1)^m \left[Y_{\ell}^{m}(\theta, \phi)\right]^*
\;,
\end{align}
where the asterisk (*) denotes complex conjugation, here and throughout this section.

\textbf{Parity Symmetry}

\begin{align}
Y_{\ell}^{m}(\pi-\theta, \phi+\pi)=(-1)^\ell Y_{\ell}^{m}(\theta, \phi)
\;.
\end{align}

\textbf{Orthogonality}
\begin{align}
\begin{split}
&
    \int_{\theta=0}^{\pi} \int_{\phi=0}^{2\pi} \left[Y_{\ell_1}^{m_1}(\theta, \phi)\right]^* Y_{\ell_2}^{m_2}(\theta, \phi) \sin \theta d \theta d \phi
\\=\;&
    \delta_{\ell_1, \ell_2} \delta_{m_1, m_2}
\end{split}
\end{align}

\subsection{Real spherical harmonics $Y_{\ell m}: S^2 \rightarrow \mathbb{R}$}
\textbf{Conventions}
\begin{align}\label{eq:SH_real_vs_complex}
Y_{\ell, m} 
&= 
\begin{cases}
    \frac{\mathrm{i}}{\sqrt{2}}(Y_{\ell}^m - (-1)^m Y_{\ell}^{-m}), & m < 0\,, \\
    Y_{\ell}^0, & m = 0\,, \\
    \frac{1}{\sqrt{2}}(Y_{\ell}^{-m} + (-1)^m Y_{\ell}^m), & m > 0 \,,
\end{cases}
\end{align}
\begin{align}
Y_{\ell}^m
&= 
\begin{cases}
\frac{1}{\sqrt{2}}\left(Y_{\ell,|m|}-\mathrm{i} Y_{\ell,-|m|}\right), &   m<0\,, \\ 
Y_{\ell, 0}, &  m=0 \,,\\ 
\frac{(-1)^m}{\sqrt{2}}\left(Y_{\ell,|m|}+\mathrm{i} Y_{\ell,-|m|}\right), &  m>0\,.
\end{cases}
\end{align}

\textbf{Orthogonality and normalization}
\begin{align}
\int_{\theta=0}^\pi \int_{\varphi=0}^{2 \pi} Y_{\ell, m} Y_{\ell^{\prime}, m^{\prime}} d \Omega=\delta_{\ell, \ell^{\prime}} \delta_{m, m^{\prime}}
\,.
\end{align}

\subsection{Basis and dual basis}

In both the physical analysis (Sec.~\ref{sec:spectral BBGKY hierarchy}) and the time evolution (Sec.~\ref{sec:Result}), we employ the basis and dual basis constructed from real spherical harmonics, as defined in Eqs.~\eqref{eq:basis} and~\eqref{eq:dual_basis}.
 
To facilitate compatibility with the use of real spherical harmonics,
we define a new set of basis and dual basis functions, denoted by $\tilde{\basis}_{n,\ell,m}$ and $\tilde{\basisleft}_{n,\ell,m}$, respectively.
These functions are defined as
\begin{align}\label{eq:basis_complex}  
\begin{split}
    \tilde{\basis}_{n,\ell,m}(p_{\mu})  
=\;&
    e^{-p_{\mu} u^\mu / \Lambda}  
    \left( \frac{p_{\mu} u^\mu}{\Lambda} \right)^\ell  
\\&\times
    Y_{\ell}^{m}(\theta,\phi)  
    L_n^{(2\ell+2)}\left( \frac{p_{\mu} u^\mu}{\Lambda} \right)  
    \,,  
\end{split}
\end{align} 
\begin{align} \label{eq:dual_basis_complex} 
\begin{split}  
    \tilde{\basisleft}_{n,\ell,m}(p_{\mu})  
=\;&
    \frac{n!}{(2\ell+n+2)!}  
    \left( \frac{p_{\mu} u^\mu}{\Lambda} \right)^{\ell+2}  
\\&\times
    Y_{\ell}^{m}(\theta,\phi)  
    L_n^{(2\ell+2)}\left( \frac{p_{\mu} u^\mu}{\Lambda} \right)  
    \,.
\end{split}  
\end{align}
Their corresponding flattened (single-index) representations are written as $\tilde{\basis}_{i}$ and $\tilde{\basisleft}_{i}$.
The basis $\basis_i$ is related to $\tilde{\basis}_i$ through a linear transformation of the form
\begin{align}
    \basis_i =\;& \mathcal{Y}_{ii'}\tilde{\basis}_{i'}\,,
\end{align}
and the corresponding dual basis $\basisleft_i$ is related to $\tilde{\basisleft}_i$ by
\begin{align}
    \basisleft_i =\;& \mathcal{Y}_{ii'}^*\tilde{\basisleft}_{i'}
    \,,
\end{align}
where the transformation coefficients $\mathcal{Y}_{ii'}$ and $\mathcal{Y}_{ii'}^*$ are defined in Eq.~\eqref{eq:SH_real_vs_complex}.

\subsection{Separated Cartesian form}

The complex spherical harmonic $Y_\ell^m(\theta,\phi)$ can be factorized into a polynomial in $z$ and another in $x$ and $y$, as follows. For $m > 0$,
\begin{align}\label{eq:ReducibleTesnor_begin}
\begin{split}
    r^{\ell}\binom{Y_{\ell}^m}{Y_{\ell}^{-m}}
=\;&
    \left[\frac{2 \ell+1}{4 \pi}\right]^{1 / 2} 
    \bar{\Pi}_{\ell}^m(z)
\\&\times
    \binom{(-1)^m\left(A_m+\mathrm{i} B_m\right)}{\left(A_m-\mathrm{i} B_m\right)}
\end{split}
\end{align}
and for $m = 0$,
\begin{align}
r^{\ell} Y_{\ell}^0 \equiv \sqrt{\frac{2 \ell+1}{4 \pi}} \bar{\Pi}_{\ell}^0 \,.
\end{align}
Here,
\begin{align}
& A_m(x, y)=\sum_{p=0}^m\binom{m}{p} x^p y^{m-p} \cos \left((m-p) \frac{\pi}{2}\right), \\
& B_m(x, y)=\sum_{p=0}^m\binom{m}{p} x^p y^{m-p} \sin \left((m-p) \frac{\pi}{2}\right),
 \end{align}
and
\begin{align}\label{eq:ReducibleTesnor_end}
\begin{split}
&
    \bar{\Pi}_{\ell}^m(z)
=\;
    \left[\frac{(\ell-m)!}{(\ell+m)!}\right]^{\frac{1}{2}} \sum_{k=0}^{\lfloor\frac{\ell-m}{2}\rfloor}(-1)^k 2^{-\ell}
\\&
    \times
    \binom{\ell}{k}\binom{2 \ell-2 k}{\ell} \frac{(\ell-2 k)!}{(\ell-2 k-m)!} r^{2 k} z^{\ell-2 k-m}\,.
\end{split}
\end{align}
\section{$3j$ symbol}\label{Appendix:3j_symbol}

The Wigner $3j$ symbol 
\begin{align*}
\left(\begin{array}{ccc}
j_1 & j_2 & j_3 \\
m_1 & m_2 & m_3
\end{array}\right)
\end{align*}
arises in the theory of angular momentum coupling and is closely related to the Clebsch--Gordan coefficients. It encodes the symmetry properties of coupled angular momentum states and appears frequently in calculations involving spherical harmonics and tensor decompositions. 

The quantities $j_1, j_2, j_3$ in the $3j$ symbol are called angular momenta. Either all of them are non-negative integers, or one is a non-negative integer and the other two are positive half-integers.
The $3j$ symbol vanishes unless two conditions are simultaneously satisfied,
(i) the triangle inequalities
\begin{align}\label{eq:3j_condition1}
\ell_1 \le \ell_2 + \ell_3, \quad \ell_2 \le \ell_1 + \ell_3, \quad \ell_3 \le \ell_1 + \ell_2
\,,
\end{align}
and
(ii)
\begin{align}\label{eq:3j_condition2}
m_1 + m_2 + m_3 = 0\,.
\end{align}
When both conditions hold, the $3j$ symbol can be evaluated using the explicit finite-sum formula provided below.
\begin{align}
\begin{aligned}
&
    \left(\begin{array}{ccc}
    j_1 & j_2 & j_3 \\
    m_1 & m_2 & m_3
    \end{array}\right)
\\=\;& 
    (-1)^{j_1-j_2-m_3} 
    \Delta\left(j_1 j_2 j_3\right)
    \sqrt{
    \left(j_1+m_1\right)!
    \left(j_1-m_1\right)!
    }
\\& \times 
    \sqrt{
    \left(j_2+m_2\right)!
    \left(j_2-m_2\right)!
    }
    \sqrt{
    \left(j_3+m_3\right)!
    \left(j_3-m_3\right)!
    }
\\& \times 
    \sum_s 
    \frac{(-1)^s}{s!}
    \frac{1}{\left(j_1+j_2-j_3-s\right)!}
\\&\quad\quad\times 
    \frac{1}{\left(j_1-m_1-s\right)!\left(j_2+m_2-s\right)!}
\\&\quad\quad\times 
    \frac{1}{\left(j_3-j_2+m_1+s\right)!\left(j_3-j_1-m_2+s\right)!}
    \,,
\end{aligned}
\end{align}
where the factor $\Delta(j_1 j_2 j_3)$ is defined by
\begin{align}
\begin{split}
&
    \Delta\left(j_1 j_2 j_3\right)
\\=\;&
    \sqrt{
    \frac{\left(j_1+j_2-j_3\right)!
    \left(j_1-j_2+j_3\right)!
    \left(-j_1+j_2+j_3\right)!
    }{\left(j_1+j_2+j_3+1\right)!}
    },
\end{split}
\end{align}
and the summation is taken over all non-negative integers $s$ such that all arguments of the factorials are non-negative.

Specifically, when $j_1 \in \mathbb{N}, j_2 \in \mathbb{N},j_3=1$, $m_1 \in \{ -j_1,\cdots,j_1\}, m_2 \in \{ -j_2,\cdots,j_2 \},m_3 \in \{ -1,0,1 \}$, and both conditions in Eqs.~\eqref{eq:3j_condition1} and \eqref{eq:3j_condition2} are satisfied, the $3j$ symbols can be written in a simple form.

\begin{align}
\begin{split}
&
    \left(\begin{array}{ccc}
    j_1 & j_2 & 1 \\
    0 & 0 & 0
    \end{array}\right)
\\=\;&
    (-1)^{j_1} 
    \left(j_1-j_2\right) 
    \left(j_1+j_2+1\right) 
\\\times&
    \sqrt{\frac{(j_1+j_2-1)!}{(j_1+j_2+2)! (j_1-j_2+1)! (-j_1+j_2+1)!}}
    \,.
\end{split}
\end{align}

\begin{align}
\begin{split}
&
    \left(\begin{array}{ccc}
    j_1 & j_2 & 1 \\
    m_1 & m_2 & 0
    \end{array}\right)
\\=\;&
    (-1)^{j_1+m_1}
    \sqrt{\frac{ (j_1-m_1)! (j_2-m_2)!}{ (j_1+m_1)! (j_2+m_2)!}}
\\\times&
    \Big(
    \left(j_1-j_2\right) \left(j_1+j_2+1\right) + \left(m_1-m_2\right)
    \Big)
\\\times&
    \sqrt{\frac{(j_1+j_2-1)!}{(j_1+j_2+2)! (j_1-j_2+1)!(-j_1+j_2+1)!}}
    \,.
\end{split}
\end{align}

\begin{align}
\begin{split}
&
    \left(\begin{array}{ccc}
    j_1 & j_2 & 1 \\
    m_1 & m_2 & 1
    \end{array}\right)
\\=\;&
    \sqrt{2} (-1)^{j_1+m_1} 
    \sqrt{\frac{(j_1-m_1)! (j_2-m_2)!}{(j_1+m_1)! (j_2+m_2)!}}
\\\times&
    \sqrt{\frac{(j_1+j_2-1)!}{(j_1+j_2+2)! (j_1-j_2+1)! (-j_1+j_2+1)!}}
    \,.
\end{split}
\end{align}

\begin{align}
\begin{split}
&
    \left(\begin{array}{ccc}
    j_1 & j_2 & 1 \\
    m_1 & m_2 & -1
    \end{array}\right)
\\=\;&
    \sqrt{2} (-1)^{j_2+m_2} 
    \sqrt{\frac{(j_1+m_1)! (j_2+m_2)!}{ (j_1-m_1)! (j_2-m_2)!}}
\\\times&
    \sqrt{\frac{(j_1+j_2-1)!}{(j_1+j_2+2)! (j_1-j_2+1)! (-j_1+j_2+1)!}}
    \,.
\end{split}
\end{align}

\textbf{Symmetry}

Even permutations of the columns of a $3j$ symbol leave it unchanged; odd permutations of the columns produce a phase factor $(-1)^{j_1+j_2+j_3}$. For example,
\begin{align}
\begin{split}
    &
    \left(\begin{array}{ccc}
    j_1 & j_2 & j_3 \\
    m_1 & m_2 & m_3
    \end{array}\right)
\\=\;&
    \left(\begin{array}{ccc}
    j_2 & j_3 & j_1 \\
    m_2 & m_3 & m_1
    \end{array}\right)=\left(\begin{array}{ccc}
    j_3 & j_1 & j_2 \\
    m_3 & m_1 & m_2
    \end{array}\right)
    \,,
\end{split}
\end{align}
\begin{align}
\begin{split}
    \left(\begin{array}{ccc}
    j_1 & j_2 & j_3 \\
    m_1 & m_2 & m_3
    \end{array}\right)
=\;&
    (-1)^{j_1+j_2+j_3}\left(\begin{array}{ccc}
    j_2 & j_1 & j_3 \\
    m_2 & m_1 & m_3
    \end{array}\right)\,,
\end{split}
\end{align}
\begin{align}
\begin{split}
    \left(\begin{array}{ccc}
    j_1 & j_2 & j_3 \\
    m_1 & m_2 & m_3
    \end{array}\right)
=\;&
    (-1)^{j_1+j_2+j_3}\left(\begin{array}{ccc}
    j_1 & j_2 & j_3 \\
    -m_1 & -m_2 & -m_3
    \end{array}\right)\,.
\end{split}
\end{align}

\section{Other useful equations}\label{Appendix:Other Physical Quantity in Coefficients}

The particle current \( J^i = \int_{\bp} p^i f(t,\bx,\bp) \) describes the flux of particles in the \( i \)-th spatial direction. When expanded in the spherical basis used in this work, each component of the current can be expressed directly in terms of the expansion coefficients \( f^{(n,1,m)} \),
\begin{align}
    J^x =\;& \sum_{n=0}^{\infty} \frac{4 \sqrt{3\pi}}{(2\pi)^3} T^3 f^{(n,1,1)}
    \,,\\
    J^y =\;& \sum_{n=0}^{\infty} \frac{4 \sqrt{3\pi}}{(2\pi)^3} T^3 f^{(n,1,-1)}
    \,,\\
    J^z =\;& \sum_{n=0}^{\infty} \frac{4 \sqrt{3\pi}}{(2\pi)^3} T^3 f^{(n,1,0)}\,.
\end{align}

The generalized Laguerre polynomials $L_n^{(2\ell+2)}(x)$ can be expressed as a finite series expansion
\begin{align}
\begin{split}
L_n^{(2\ell+2)}(x) 
& =
\sum_{f=0}^n(-1)^f\binom{n+2\ell+2}{n-f} \frac{x^f}{f!}
\end{split}\label{eq:laguerrepolynomials}
\end{align}

\newpage
\bibliography{biblio.bib}
\end{document}